\documentclass{article}
\pdfoutput=1

\usepackage[top=0.85in,left=1.75in,right=1.75in,footskip=0.75in]{geometry}

\usepackage{amsmath,amssymb}
\usepackage{changepage}
\usepackage[utf8x]{inputenc}
\usepackage{textcomp,marvosym}
\usepackage{cite}
\usepackage{nameref,hyperref}
\usepackage{microtype}
\DisableLigatures[f]{encoding = *, family = * }
\usepackage[table]{xcolor}
\usepackage{array}
\newcolumntype{+}{!{\vrule width 2pt}}
\newlength\savedwidth

\usepackage{setspace} 

\usepackage{lastpage,fancyhdr,graphicx}

\title{Addressing nonlinearities in Monte Carlo}

\author{J\'er\'emi Dauchet$^{1}$ \and  Jean-Jacques Bezian$^{2}$ \and  St\'ephane Blanco$^{3}$ \and  Cyril Caliot$^{4}$ \and Julien Charon$^{1}$ \and  Christophe Coustet$^{5}$ \and  Mouna El Hafi$^{2}$ \and  Vincent Eymet$^{5}$ \and  Olivier Farges$^{6}$ \and  Vincent Forest$^{5}$ \and  Richard Fournier$^{3}$ \and  Mathieu Galtier$^{7}$ \and  Jacques Gautrais$^{8}$ \and  Ana\"is Khuong$^{8}$ \and  Lionel Pelissier$^{9}$ \and  Benjamin Piaud$^{5}$ \and  Maxime Roger$^{7}$ \and  Guillaume Terr\'ee$^{2}$  \and  Sebastian Weitz$^{1}$}
\date{}

\begin{document}
\maketitle

\begin{enumerate}
 \item Universit\'e Clermont Auvergne, Sigma-Clermont, Institut Pascal, BP 10448, F-63000 Clermont-Ferrand, France.
 \item Universit\'e F\'ed\'erale de Toulouse Midi-Pyr\'en\'ees, Mines Albi, UMR CNRS 5302, Centre RAPSODEE, Campus Jarlard, F-81013 Albi CT Cedex.
 \item LAPLACE, Universit\'e de Toulouse, CNRS, INPT, UPS, France.
 \item Processes, Materials and Solar Energy Laboratory, PROMES, CNRS, 7 rue du Four Solaire, 66120 Font-Romeu-Odeillo-Via, France.
 \item M\'eso-Star SAS, www.meso-star.com.
 \item Total Energies Nouvelles, R\&D - Concentrated Solar Technologies, Tour Michelet, 24
cours Michelet, 92069 Paris La D\'efense Cedex, France.
 \item Univ Lyon, CNRS, INSA-Lyon, Universit\'e Claude Bernard Lyon 1, CETHIL UMR5008, F-69621, Villeurbanne, France.
 \item Research Center on Animal Cognition (CRCA), Center for Integrative Biology (CBI), Toulouse University, CNRS, UPS, France.
 \item UMR MA 122 EFTS (Education, Formation, Travail, Savoirs), Universit\'e Toulouse 2, Toulouse.
\end{enumerate}

\clearpage

\begin{abstract}
{
Monte Carlo is famous for accepting model extensions and model refinements up to infinite dimension. However, this powerful incremental design is based on a premise which has severely limited its application so far: a state-variable can only be recursively defined as a function of underlying state-variables if this function is linear. Here we show that this premise can be alleviated by projecting nonlinearities onto a polynomial basis and increasing the configuration space dimension. Considering phytoplankton growth in light-limited environments, radiative transfer in planetary atmospheres, electromagnetic scattering by particles, and concentrated solar power plant production, we prove the real-world usability of this advance in four test cases which were previously regarded as impracticable using Monte Carlo approaches. We also illustrate an outstanding feature of our method when applied to acute problems with interacting particles: handling rare events is now straightforward. Overall, our extension preserves the features that made the method popular: addressing nonlinearities does not compromise on model refinement or system complexity, and convergence rates remain independent of dimension.

Published: Dauchet J, Bezian J-J, Blanco S, Caliot C, Charon J, Coustet C, El Hafi M, Eymet V, Farges O, Forest V, Fournier R, Galtier M, Gautrais J, Khuong A, Pelissier L, Piaud B, Roger M, Terr\'ee G, Weitz S (2018) Addressing nonlinearities in Monte Carlo. Sci. Rep. 8: 13302, DOI:10.1038/s41598-018-31574-4
}
\end{abstract}

\clearpage

\section{Introduction}
The standard Monte Carlo (MC) method is a technique to predict a physical observable by numerically estimating a statistical expectation over a multi-dimensional configuration space\cite{Dimov08}.
The reason why this method is so popular in all fields of scientific research is its intuitive nature. 
In general, simulation tools are designed in direct relation to the physical phenomena present in each discipline, and later refinements are gradual and straightforward. 
Model refinements merely extend sampling to other appropriate dimensions. 
The method is nonetheless mathematically rigorous: specialists specify observables that are implicitly translated into integral quantities which are estimated using random sampling in each direction of the configuration space.
This statistical approach is highly powerful because the algorithm can be designed directly from the description of the system, whether it is deterministic or not, with no reworking or approximation.

Let us illustrate how MC is used in engineering with a typical example: the optimal design of a concentrated solar plant\cite{Delatorre13} (see Fig.1-a).
The power collected by the central receiver results from all the rays of sunlight that reach it after reflection by heliostats, so it depends on the complex geometry of the heliostats.
Moreover, the heliostats change their orientation to follow the sun’s position, so they can mask one another at certain times of the day. 
To estimate by MC the received power at one moment of interest, i.e. for a given geometry of the heliostats: choose an optical path among those that link the sun to the central receiver via a heliostat; check whether this path is shadowed or blocked by another heliostat; and retain a Monte Carlo weight equal to $0$ or $1$ depending on transmission success. 
Let $\bf X$ be the random variable denoting transmission success.
The collected fraction of the available sun power is then the expectation ${\cal E}_{\bf X}(\bf X)$ of $\bf X$, and can be evaluated with no bias as the average of such weights over a large number of sampled paths. 

\bigskip
This approach robustly complies with expanded descriptions of the physical observable to be addressed. 
For instance, the fraction of the available sun power collected on average over the entire lifetime of the solar plant (typically 30 years) can be predicted as the expectation over time of ${\cal E}_{\bf X}(\bf X)$, which varies with time. Denoting ${\cal E}_{\bf X|Y}(\bf X|Y)$ the collected fraction at random time $\bf Y$ within the 30 years, the time-averaged fraction is given by ${\cal E}_{\bf Y}({\cal E}_{\bf X|Y}({\bf X|Y}))={\cal E}_{{\bf Y},{\bf X|Y}}({\bf X|Y})$.
The basic algorithm above can then be encapsulated within time sampling: first choose a date for $\bf Y$, then pick a path at that date for $\bf X|Y$. 
Finally, estimate ${\cal E}_{{\bf Y},{\bf X|Y}}({\bf X|Y})$ by computing the average transmission success over all combined pairs (date, path).
Meanwhile, sun power fluctuations can be accounted for by estimating the atmospheric transmission at each chosen date.
The choice of the statistical viewpoint thus enables us to incorporate into one single statistical question as many elements as necessary: the geometrical complexity of the heliostats\cite{Siala01}, the daily course of the sun, and seasonal-scale as well as hourly-scale weather fluctuations\cite{Farges15}. 
Remarkably, the latter question is nearly as simple to address as the estimation of the power collected at one single date: the algorithmic design can map the full conceptual description, yet computational costs are hardly affected.
Contrastingly, deterministic approaches would translate into impractical computation times or require simplified and approximate descriptions, so MC has become the only practical solution in many engineering contexts of this type. 
Having become standard practice, MC has prompted numerous theoretical developments\cite{Metropolis53,Hammersley80,Assaraf99,Roger05}.

\clearpage
Figure 1 --- \textbf{Complex systems with nonlinear outputs: four real-world examples.}
\vspace{0.5cm}

\includegraphics[width=13.5cm]{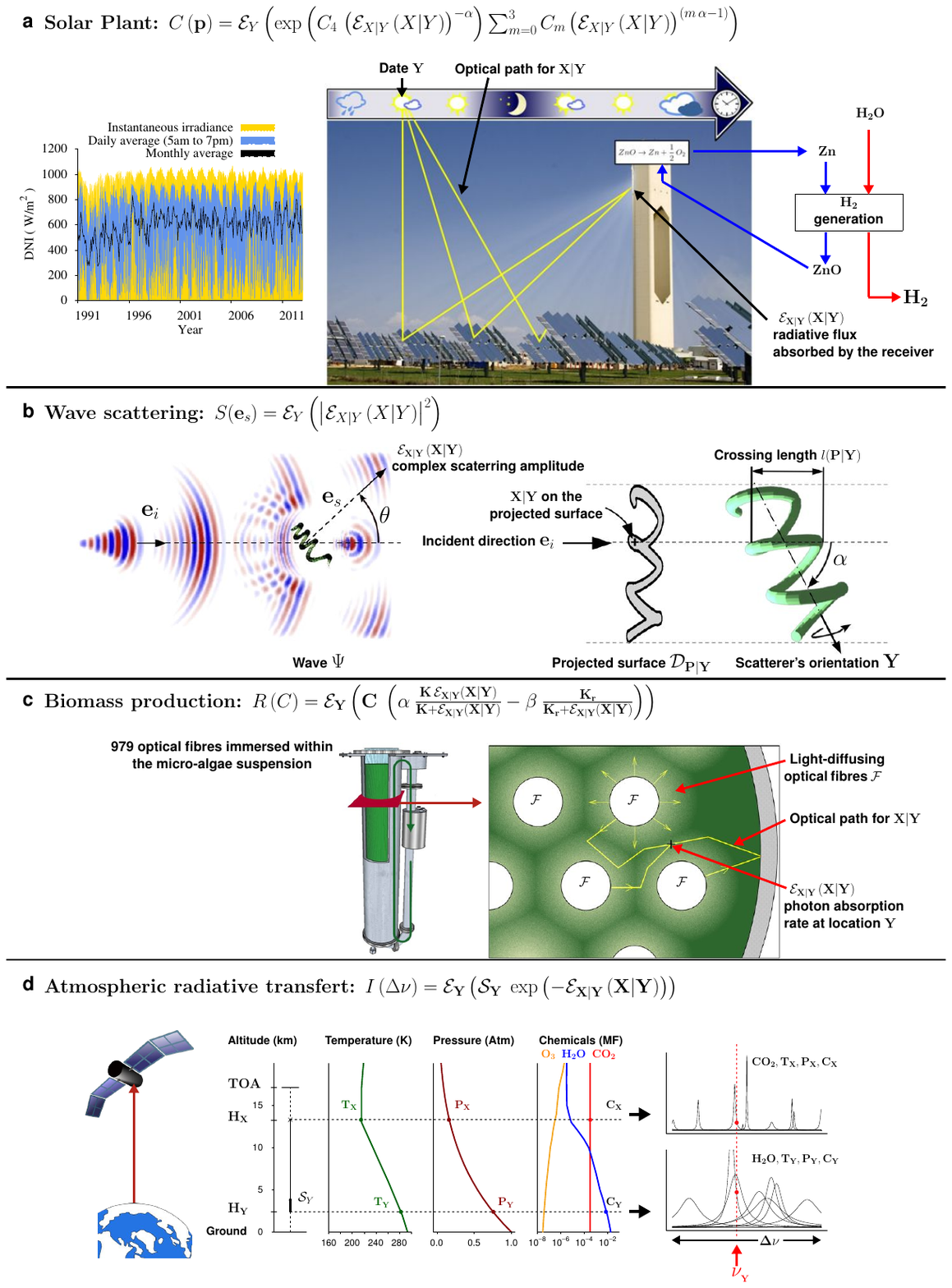}
\clearpage
\paragraph{Fig. 1 \textbf{Complex systems with nonlinear outputs: four real-world examples.}}
$\;$\\
\textbf{a, Solar-driven high-temperature thermal reduction of zinc oxide, as the first phase of a two-step water-splitting cycle.} 
Photons emitted from the sun are reflected on heliostats and concentrated at the entrance of the chemical reactor in which $ZnO$ dissociation is carried out. Depending on their transmission success $\bf X$, the solar power ${\cal E}_{\bf X|\bf Y} \left( {\bf X|\bf Y}\right)$ absorbed by the receiver at a random instant ${\bf Y}$ determines the nonlinear chemical conversion rate of the reaction $ZnO \rightarrow Zn + \frac{1}{2} O$. 
Here we address the estimation of the solar plant's annual conversion rate $C(\bf p)$ at different Earth locations $\bf p$, by averaging the instantaneous conversion rates over the statistics of sun position and incident Direct Normal Irradiance (DNI), which fluctuates with time and weather conditions (see also SI1).
$\;$\\
\textbf{b, Wave scattering by a complex-shaped and optically soft scatterer (cyanobacterium {\it Arthrospira}).}
An incident plane wave with propagation direction ${\bf e}_i$ is scattered by the helical cyanobacterium. The bacterium has low relative refractive index and is much larger than the wavelength (optically soft particle).
The complex scattering amplitude ${\cal E}_{\bf X|\bf Y} \left( {\bf X|\bf Y}\right)$ in the forward directions is the sum of secondary wave contributions ${\bf X|\bf Y}$ (interference) originating from the scatterer surface. This surface depends on the scatterer orientation ${\bf Y}$. Here we address the estimation of $S({\bf e}_s)$, the single-scattering differential cross-section, in direction ${\bf e}_s$ for a suspension of particles, assuming independent scattering, by averaging the squared modulus of ${\cal E}_{\bf X|\bf Y} \left( {\bf X|\bf Y}\right)$ over the statistics of orientations ${\bf Y}$ (see also SI2).
$\;$\\
\textbf{c, Phytoplankton growth in light-limited environments.} 
Phytoplankton is grown in a continuous stirred tank photobioreactor, internally illuminated by optical fibres ${\cal F}$ immersed in the culture.
The local rate of photon absorption ${\cal E}_{\bf X|\bf Y} \left( {\bf X|\bf Y}\right)$ at location $\bf Y$ is the average of the contributions ${\bf X}|{\bf Y}$ of every optical path from the fibres to $\bf Y$ through the scattering and absorbing suspension. ${\cal E}_{\bf X|\bf Y} \left( {\bf X|\bf Y}\right)$ determines the nonlinear photosynthetic growth rate at location ${\bf Y}$.
Here we address the Monte Carlo estimation of $R(C)$, the global growth-rate in the whole culture volume, as a function of biomass concentration $C$, by averaging the local rate over locations in the volume (see also SI3). 
$\;$\\
\textbf{d, Atmospheric radiative transfer : top-of-atmosphere (TOA) specific intensity (from earth towards outer space).}
Photons emitted by the atmosphere at infrared frequencies are due to random emission transitions ${\bf Y}$, from a higher to a lower energy state, of mainly $\mathrm{CO_2}$ and $\mathrm{H_2O}$ molecules of concentration $C_{\bf Y}$ at altitude $H_{\bf Y}$. The corresponding source ${\cal S}_{\bf Y}$ depends on the thermodynamic state of the atmosphere, mainly temperature $T_{\bf Y}$ (defining the energy-state population) and pressure $P_{\bf Y}$ (defining most of the line width, i.e. the uncertainty of the emission frequency $\nu_{\bf Y}$). This source is then exponentially attenuated by atmospheric absorption, i.e. by all random absorption transitions ${\bf X}|{\bf Y}$, from a lower to a higher energy state, occurring at altitude $H_{\bf X|Y}$ between $H_{\bf Y}$ and the top of the atmosphere (see also SI4).\\
\textbf{Copyright.} "Central solar PS10" (https://commons.wikimedia.org/wiki/File:Luz.jpg) by MwAce is released into the public domain. "satellite" (www.pixabay.com/en/satellite-solar-panels-space-297840), "blue-earth" (pixabay.com/en/earth-blue-land-globe-planet-297125), "forecast icons" (www.pixabay.com/en/weather-signs-symbols-forecast-28719) and "time-clock" (pixabay.com/en/clock-time-hour-minute-wall-clock-295201) by Clker-Free-Vector-Images (www.pixabay.com/en/users/Clker-Free-Vector-Images-3736) are licensed under CC0 Creative Commons (www.creativecommons.org/publicdomain/zero/1.0/deed.en).
\clearpage

Nevertheless, MC has so far not been able to handle \emph{every} question. 
In fact, it was identified early on that ``the extension of Monte Carlo methods to nonlinear processes may be impossible''\cite{Curtiss54} and it is a prevalent opinion nowadays that ``Monte Carlo methods are not generally effective for nonlinear problems, mainly because expectations are linear in character''\cite{Kalos08}, so that ``a nonlinear problem must usually be linearized in order to use the Monte Carlo technique''\cite{Chatterjee14}.
We are aware of only one attempt so far to bypass this failing: the recent proposal by the applied mathematics community\cite{Dimov08,Vajargah07,rasulov10,gobet16} to use branching processes\cite{skorokhod64} to solve Fredholm-type integral equations with polynomial nonlinearity.

Unfortunately, most real-world problems are nonlinear. 
Indeed, if the question were now to evaluate the final return on investment of the solar plant, namely how much electrical power it would deliver over its lifetime, standard MC would fail, because the instantaneous conversion efficiency from collected solar power to electrical power is not linear. 
Let us consider, as a toy example, a basic nonlinear case where the electrical power would be proportional to the square of the instantaneous collected solar power ${\cal E}_{\bf X|Y}(\bf X|Y)$ at date $\bf Y$. In Monte-Carlo terms, the question would then be to estimate ${\cal E}_{\bf Y}({\cal E}_{\bf X|Y}({\bf X|Y})^{2})$ over the plant’s lifetime.
In this case, the optical and temporal expectations can no longer be combined, because it would be wrong to first estimate, as above, the total solar power collected over its lifetime, and then apply the conversion efficiency at the end (basically, ${\cal E}_{\bf Y}({\cal E}_{\bf X|Y}({\bf X|Y})^{2}) \ne {\cal E}_{\bf Y}({\cal E}_{\bf X|Y}({\bf X|Y}))^{2}$, in the same way as $a^{2}+b^{2} \ne (a+b)^{2}$). 
Instead, we would have to sample dates (say $M$ dates, millions over 30 years), estimate the solar power collected at each date by averaging transmission successes over numerous optical paths (say $N$ paths, millions for each date), apply a nonlinear conversion to the result at that date, and then average over all dates\cite{NLexpectation}. 
Doing so, MC would now require $M\times N$ samples, and even worse, further levels of complexity (each adding a nonlinearity to the problem) would similarly multiply the computation time.
Moreover, the result would be biased due to the finite sampling sizes of the innermost dimensions.
In short, MC’s distinctive features are no longer available, and exact lifetime integration appears impossible.

\section{Results}
Bearing in mind our earlier theoretical works about MC integral formulations\cite{Delatorre13}, we have found a way to bypass this obstacle for a large class of nonlinear problems, based on the very statistical nature of MC.
In the case of our toy example, we use the fact that:
\begin{equation}
{\cal E}_{\bf Y}({\cal E}_{\bf X|Y}({\bf X|Y})^{2})={\cal E}_{{\bf Y},{\bf (X_1,\,X_2)|Y}}({\bf X_{1}\,X_{2}|Y})
\end{equation}
where ${\bf X}_{1}$ and ${\bf X}_{2}$ are two independent variables, identically distributed as ${\bf X}$ (see Methods).
Translated into a sampling algorithm, the solution is now to sample optical paths in pairs ${\bf (X_1,X_2)|Y}$ (instead of millions) at each sampled date, and then to retain the pair product ${\bf X_{1}X_{2}|Y}$ of their transmission successes.
The optical and temporal statistics can then actually be sampled together, and yield the unbiased result with no combinatorial explosion.
This reformulation can be generalised to any nonlinearity of polynomial shape.
First, monomials of any degree can indeed be estimated using the same sampling property as that used above for $n=2$:
\begin{equation}
{\cal E}_{\bf Y}({\cal E}_{\bf X|Y}({\bf X|Y})^{n})={\cal E}_{{\bf Y},{\bf (X_1,\,X_2, ...,\,X_n)|Y}}({\bf X_{1}\,X_{2} ...X_{n}|Y})
\end{equation}
where ${\bf X}_i$ are $n$ independent random variables, identically distributed as ${\bf X}$. 
For any monomial of degree $n$, the expectation can then be computed by sampling series of $n$ independent realisations of $\bf X|Y$, and averaging the series products.
The linear case, solved by standard MC, corresponds to $n=1$.
Secondly, since polynomials are simply linear combinations of monomials, the expectation for any polynomial function of ${\cal E}_{\bf X|Y}({\bf X|Y})$ of degree $n$ can be translated into a Monte Carlo algorithm, first sampling a degree in the polynomial, and then sampling as many independent realisations of ${\bf X|Y}$ as this random degree (see {\it Methods}). 
For a polynomial function of degree $n$, the corresponding Non-Linear Monte Carlo (NLMC) algorithm is then:
\begin{itemize}
\item pick a sample $y$ of ${\bf Y}$,
\item choose a monomial degree value $d \le n$,
\item draw $d$ independent samples of ${\bf X|Y=y}$ and retain their product,
\end{itemize}
repeat this sampling procedure and compute the estimate as the average of the retained products.

Moreover, if polynomial forms of any dimension are now solvable with no approximation, so is the projection of any nonlinear function onto a polynomial basis of any dimension, even of infinite dimension if required (full details of using the Taylor expansion are given in {\it Methods}).
As a result, any hierarchy of nested statistical processes that combine nonlinearly can now, in theory, be exactly addressed within the Monte Carlo framework. The deep rationale of the proposed algorithm is therefore to transform a nonlinear process into a formally equivalent linear infinite-dimension process, and then use the inherent capability of Monte Carlo to address expectations over domains of infinite dimension.

To the best of our knowledge, this analysis has never before been performed. However, it has major practical consequences for real-world problems, provided the polynomial sampling, which is the price to be paid for tackling nonlinearities exactly, remains tractable.
For instance, let us go back to our solar power plant example, and now use the actual expression for the conversion rate and its Taylor expansion: for each date, once a sun position and climate conditions have been fixed, we would have to pick a random number of independent optical paths (instead of one optical path in the linear case), keep the product of their transmission success, and finally calculate the average of many such products.
Doing so, it becomes possible to integrate hourly solar input fluctuations over 30 years in the full geometry of a kilometre-wide heliostat field in order to optimise the nonlinear solar-to-electric conversion over the plant lifetime (Fig. 1A).
The same line of thought can be used to predict wave scattering by a tiny complex-shaped scatterer\cite{Charon_16} such as a helicoidal cyanobacterium (Fig. 1B). 
The biomass production example (Fig. 1C), where incoming light favours the photosynthetic growth that in turn blocks the incoming light, illustrates how our method also handles nonlinear feedback\cite{Cornet2010}.
Finally, with the estimation of Earth’s radiative cooling (Fig. 1D), we reproduce quite classic results, yet with a purely statistical approach: by sampling directly the state transitions of greenhouse gases, we avoid costly deterministic computations that the standard linear Monte Carlo approach requires in order to by-pass the nonlinearity of the Beer Extinction Law\cite{Galtier2015}.
In each of the four cases, it appears that the additional computations are well affordable using only ordinary computing power 
(the complete physical descriptions of the four problems, the nonlinearities involved and their translation in NLMC can be found in their respective Extended Data Figures and Supplemental Information, Solar Plant: SI1 ; Complex-shaped Scatterer: SI2 ; Biomass production: SI3 ; Earth radiative cooling: SI4).

For these four real-world simulation examples, we can therefore retain that the variance of the proposed statistical estimate was very much satisfactory. Is that a general feature? Can we feel confident when applying this simulation strategy to any new nonlinear problem? More generally speaking, what do we claim about the status of the present research? Essentially, we only argue that the general proposition of the present paper is immediately available for an ensemble of pratical applications. Indeed, these four simulation examples are representative of a quite wide ensemble of physics/engineering practices and the corresponding implementations are now practically used by the corresponding research-communities\cite{Charon_16,Galtier2015,Dauchet_14,Dauchet2016}. Moreover, implementation only required an up to date knowledge of Monte Carlo techniques: the probability sets were selected using nothing more than very standard importance-sampling reasoning (see Methods, SI1 and SI3). Outside these experiments, we did not explore in any systematic manner the statistical convergence difficulties that could be specifically associated to the proposition. But although we did not yet encounter it, we can already point out a potential source of variance related to the choice of the fixed point $x_0$ around which the nonlinear function is Taylor expanded (see Methods).

From a theoretical point of view, in the four cases exposed above, the model is directly enunciated in statistical terms, defining two random variables ${\bf X}$ and ${\bf Y}$ from the start.
More broadly, standard MC practice can also start from a deterministic description (see Methods), most commonly from a linear partial differential equation (PDE).
The formal equivalence between the solution of a linear PDE and the expectation of a random variable has long been established \cite{FeynmanKac}. Indeed, PDE-to-MC translations are essential to nanoscale mechanics (Quantum Monte Carlo\cite{Corney04}) or nuclear sciences.
NLMC allows such translations for nonlinear PDEs.
 
As an illustration of the ground-breaking nature of our study, we address a prominent example of a nonlinear PDE in statistical physics, the Boltzmann equation, which governs the spatiotemporal density of interacting particles in a dilute gas (full details in SI5). 
The corresponding physics is easy to visualise: a particle simply follows its ballistic flight until it collides with another particle.
The collisions are considered as instantaneous and only modify the two particle velocities.
The equation for the variation in particle density in phase-space (position, velocity) is nonlinear because the collision rate depends on the density itself.
In order to project this nonlinearity onto the proper polynomial basis of infinite dimension, this PDE is first translated into its Fredholm integral counterpart (a step reminiscent of the aforementioned Dimov proposition\cite{Dimov08}).
This Fredholm integral expresses the density in phase space at some location for some velocity at some time, as if putting a probe into space-time.
It is estimated by Monte Carlo, tracking the dynamics backwards in time up to the initial condition (or boundary conditions). 
Importantly, such a probe estimation does not require the exhaustive resolution of the whole field at previous times: as in standard backwards MC algorithms for solving linear transport (e.g. simulating an image by tracking photon-paths backward, from receiver to source\cite{pbrt,Case1957,Collins1972}) the information about previous states of the field is reconstructed along each path only where and when it is required\cite{Galtier13}. Here, the contrast with linear MC is that nonlinearity due to collisions translates into branching paths.

\bigskip
This extension deals very efficiently with extremely rare events because it preserves an essential feature of MC: by avoiding time / space / velocity discretisation\cite{Wagner1995,Rjasanow1996,Rjasanow01}, very low densities can be estimated with no bias, and the only source of uncertainty is the finite number of sampled events (i.e. the confidence interval around the estimated density).
As a test, we consider a case for which analytical solutions have been published: Krook's early analysis of the distribution of speeds in extremely out-of-equilibrium conditions\cite{Krook76,Krook77}.
Krook's analysis was outstanding because it provided an analytical solution to a problem which looked impossible to solve numerically: events with the greatest consequences, namely the particles with the highest energies (i.e. high-speed particles, of tremendous importance in nuclear chemistry) lie far out in the tail of speed distribution and have a very low probability of occurrence (rare events). 
Using our NLMC design, the fractions of particles which have a kinetic energy higher than $10^{6}$ times the average value, and which correspond to a fraction as low as $10^{-11}$ of the total, can now be quantified as accurately as desired, and perfectly fit the analytical solution (fig. 2a).

Having been validated in Krook's case, this extension opens the way to solving systems for which no analytical solutions are available.
As an example, we now consider a fully spatialised system in which the particles are confined by an outside harmonic potential, leading to a so-called \emph{breathing mode} of the gas density. 
Such a solution to the Boltzmann equation was identified early on by Boltzmann himself\cite{BoltzmannDGO}, but has recently been revisited and generalised in the context of a shortcut to adiabaticity techniques for classic gases\cite{gdo14}.
Exact solutions are available only under the constraint that the gas is at local equilibrium, in which case the density displays a permanent oscillation.
Here again, these analytical solutions are exactly recovered.
Moreover, NLMC enables us to go beyond this constraint and to explore the gas behaviour when the local equilibrium constraint is alleviated: starting from a state far from local equilibrium, it is now possible to estimate how fast the velocity redistribution induced by collisions actually dampens the oscillation (fig. 2b).

\clearpage
\begin{center}
Figure 2 --- \textbf{Nonlinear Monte Carlo for gas kinetics.}

\vspace{0.5cm}

\includegraphics[width=14cm]{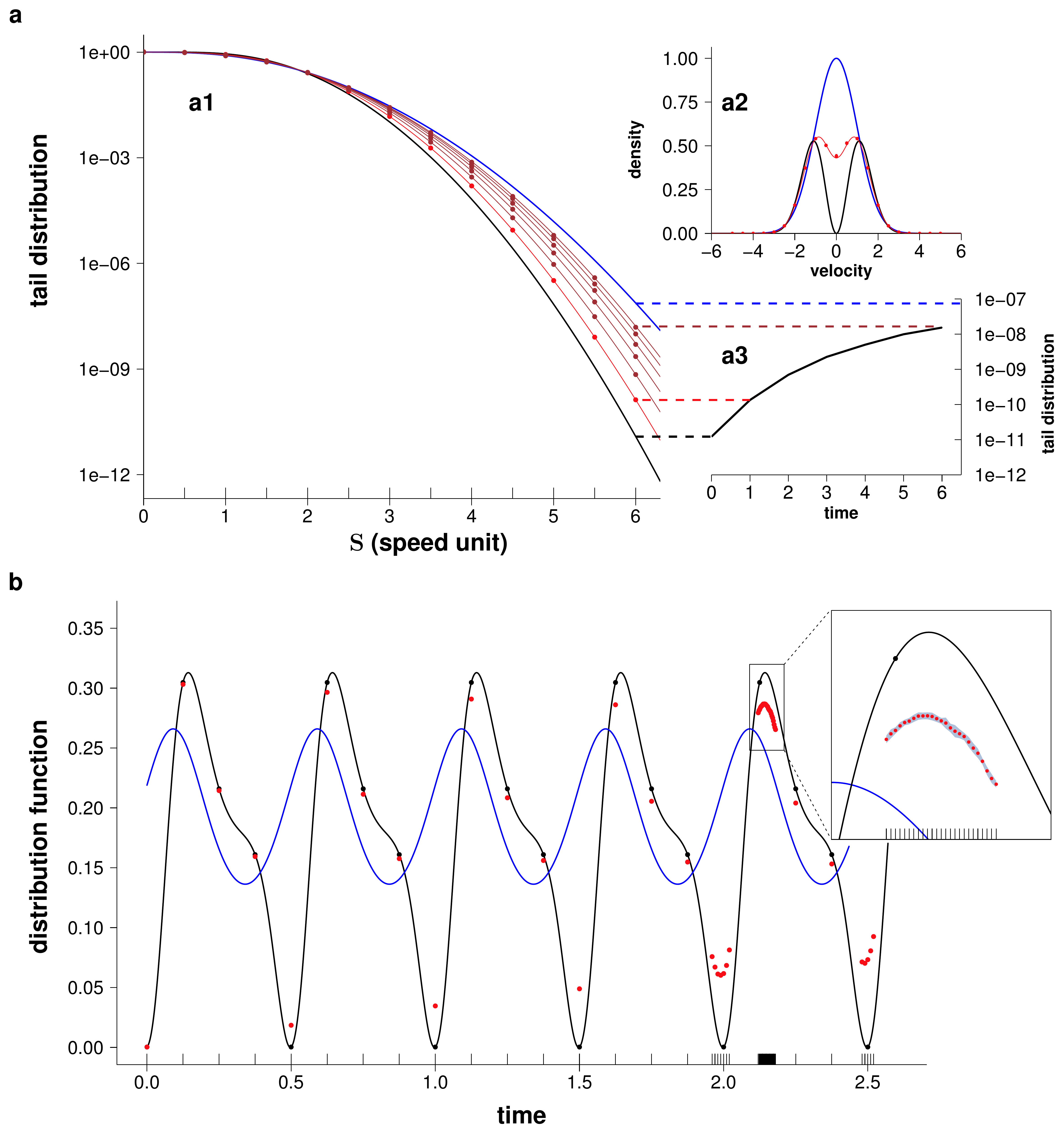}

\end{center}
\clearpage

\paragraph{Fig. 2 \textbf{Nonlinear Monte Carlo for gas kinetics.}}
$\;$\\
\textbf{a, Relaxation of speed distribution to equilibrium}
(\textbf{a1}: tail distribution of particle speeds (fraction of particles faster than $\bf S$),
\textbf{a2}: probability density function of velocity, \textbf{a3}: variation in the fraction of particles faster than 6 speed units). 
In a homogenous gas, collisions between particles redistribute velocities so that the speed distribution tends to equilibrium (Maxwellian distribution).
Starting from a distribution far from equilibrium (black curves), we compute the relaxation to a Maxwellian distribution (blue curves) by estimating the tail distribution at different times (e.g. red curves correspond to the system state at 1 unit time).
The continuous lines correspond to the analytical solutions
and each point corresponds to an independent NLMC estimation.   
The fraction of particles which are faster than 6 speed units (\textbf{a3}) illustrates how NLMC well accounts for the 1000-fold increase in the rarest high-speed particles, with no space or time discretisation. Remarkably, rare events are estimated with the same relative uncertainty as frequent events ($10^{4}$ samples for each estimate; confidence intervals of all estimates are contained within the thickness of the point). 
$\;$\\
\textbf{b, Dampening of breathing mode.}
A dilute gas confined by an outside harmonic potential displays a \emph{breathing mode}.
We estimate the density at probe position $(1.75,0,0)$ and velocity $(0.35,0,0)$ (in adimensional units, see SI5) at different times (adim. unit).
Starting from a distribution complying with the local equilibrium, the density displays a perpetual oscillation at twice the trap frequency (blue curve, analytical solution), independent of the collision rate (or, equivalently, of the elastic cross section).
Starting from a distribution which is far from the local equilibrium (the same initial distribution as in \textbf{a}), the density still pulsates in the absence of collisions (null cross section; black curve: analytical solution; black points: probe estimates). 
When collisions are introduced (by raising the cross section), the velocity redistribution induced by the collisions dampens the oscillation (red points: probe estimates; no analytical solution).
The inset illustrates that probes can be bunched together to zoom in on a period of special interest (e.g. estimating the peak values at each cycle). 
Each point is estimated independently using $10^{7}$ samples; confidence intervals of all estimates are contained within the thickness of one point in the main figure, and confidence intervals are represented by the grey background area in the inset). 
\clearpage

\section{Conclusions}
From now on, the Monte Carlo Method is no longer restricted to linear problems.
The five examples exposed above were worked out by teams comprising specialists of the Monte Carlo method and specialists of the physical problem under consideration. 
Through their complete description, we offer readers all the details to implement their own applications.
As a guideline, the first step is to formulate the physical observable under its expectation form, including the nonlinearities and integrating all levels of complexity. 
The second step is to reformulate this expectation as a formulation compliant with the standard Monte Carlo Method, according to the type of nonlinearity.
For polynomial nonlinearities, use i.i.d. series products.
For other differentiable forms, use a Taylor expansion around an upper bound of the innermost random variable in order to regain a polynomial form.
Using this MC-compliant formulation, every advanced MC technique can then be applied: parallel implementation, complex geometry, null collisions, zero variance, control variate, importance sampling, sensitivity analysis, and so on.
As illustrated by the variety of our seminal examples, this guideline covers a large set of nonlinear academic and real-world problems.

\clearpage

\begin{description}
 \item[Competing Interests] The authors declare that they have no
competing interests.
 \item[Correspondence] Correspondence and requests for materials
should be addressed to Jérémi Dauchet~(email: jeremi.dauchet@sigma-clermont.fr).
 \item[Authors contributions] All authors contributed extensively to the theoretical developments presented in this paper.
Each author contributed to the practical applications according to his or her scientific expertise: J.D., M.E.H., V.E., R.F. and M.G. in Atmospheric sciences, J.D., S.B., C.Ca., M.E.H., V.E., O.F., R.F., M.G. and M.R. in Radiative Transfer, J.D., S.B., R.F., J.G., A.K. and S.W. in Complex Systems in Biology, J.D., J.J.B., S.B., C.Ca., M.E.H., V.E., O.F. and R.F. in Solar Energy, J.D., S.B., J.C., M.E.H. and R.F. in Electromagnetism and quantum mechanics, J.D., S.B., M.E.H., R.F., J.G., B.P. and G.T. in Statistical Physics.
C.Co., V.E., V.F. and B.P. (www.meso-star.com) performed the numerical implementations. 
J.D., S.B., R.F. and J.G. wrote the paper.
 \item[Acknowledgements] This work was sponsored by the French National Centre for Scientific Research through the PEPS-JCJC OPTISOL-mu program, and by the French government research-program "Investissements d'avenir" through the LABEXs ANR-10-LABX-16-01 IMobS3 and ANR-10-LABX-22-01 SOLSTICE and the ATS program ALGUE of IDEX ANR-11-IDEX-02 UNITI. The authors express their deep gratitude to Igor Roffiac for fruitful discussions on the statistical engineering of complex systems.
\end{description}

\clearpage

\section*{Methods}
\paragraph{Basics of Monte Carlo Methods.}
Let us estimate $E=1+4$ by repeatedly tossing a (fair) coin. 
The tossing process is described by a random variable $R \in \{0,1\}$ which takes the value $0$ for heads (probability $P_R(0)=\frac{1}{2}$) and $1$ for tails (probability $P_R(1)=\frac{1}{2}$).

Now, to estimate any process (e.g. a process output: $E=1+4$), we can assign arbitrary weights $w(R)$ to values $\{0,1\}$ in order to write $E$ as an expectation of the weighted process, following: 
\begin{equation}
E=1+4 = P_R(0) w(0) + P_R(1) w(1) = {\cal E}_R \left(w(R)\right) 
\label{eq:somme}
\end{equation}
with $w(0)=\frac{1}{P_R(0)}=2$ and $w(1)=\frac{4}{P_R(1)}=8$ and where ${\cal E}_R$ denotes the expectation with respect to $R$.
Using the results $r_1 ... r_N$ of $N$ successive tosses (independent realisations of $R$), we can then estimate $E= {\cal E}_R \left(w(R)\right)$ from the weighted average of the toss results
$\frac{1}{N}\sum_{i=1}^{N}w(r_i)$ since $E=5$ is indeed the average of Monte Carlo weights that take the values $2$ and $8$ with equal probabilities.

Such an approach is at the base of Monte Carlo techniques: define the weights according to the problem to be solved, sample the process repeatedly, and take the average. 
Depending on the physical description of the value to be estimated, this approach still holds for an infinite number of terms and can also be extended to integral formulation using continuous random variables:
\begin{equation}
{\cal E}_{\bf Y} \left(w({\bf Y})\right) = \int_{{\cal D}_{\bf Y}}d{\bf y}\,p_{\bf Y}({\bf y})\,w({\bf y})
\label{eq:defMMC}
\end{equation}
which can be estimated by $\frac{1}{N}\sum_{i=1}^{N}w({\bf y}_i)$, where the $y_i$ are $N$ realisations of the random variable $Y$ with probability density function $p_Y$ and domain of definition ${\cal D}_{\bf Y}$.
\paragraph{Basics of Nonlinear Monte Carlo Methods.}

In order to estimate
\begin{equation}
E={\cal E}_{\bf Y}({\cal E}_{\bf X|Y}({\bf X|Y})^{2}) = \int_{{\cal D}_{\bf Y}} d{\bf y}\, p_{\bf Y}(y) \left( \int_{{\cal D}_{\bf X|Y}} dx\, p_{\bf X|Y}(x|y)\; x \right)^{2}
\end{equation}
we introduce two independent variables ${\bf X}_{1}$ and ${\bf X}_{2}$, identically distributed as ${\bf X}$ (still conditioned by the same ${\bf Y}$):
\begin{equation}
\begin{array}{lcl}
E&=&{\cal E}_{\bf Y}\left( {\cal E}_{\bf X_1|Y}({\bf X_{1}|Y})\,{\cal E}_{\bf X_2|Y}({\bf X_{2}|Y})\right)\\
&=&\displaystyle \int_{{\cal D}_{\bf Y}} dy\,p_{\bf Y}(y) \left( \int_{{\cal D}_{\bf X|Y}} dx_{1}\,p_{\bf X|Y}(x_{1}|y)\; x_{1} \right)\left( \int_{{\cal D}_{\bf X|Y}} dx_{2}\,p_{\bf X|Y}(x_{2}|y)\; x_{2} \right)
\end{array}
\end{equation}
Since ${\bf X}_{1}$ and ${\bf X}_{2}$ are independent, and conditionally independent given $\bf Y$:
\begin{equation}
\begin{array}{lcl}
E&=&\displaystyle \int_{{\cal D}_{\bf Y}} dy\,p_{\bf Y}(y) \left( \iint_{{\cal D}^2_{\bf X|Y}} dx_{1}\,p_{\bf X|Y}(x_{1}|y)\; dx_{2}\, p_{\bf X|Y}(x_{2}|y) \;x_{1}x_{2}  \right)\\
&=&{\cal E}_{\bf Y}\left( {\cal E}_{\bf (X_1,\,X_2)|Y}({\bf X_{1}X_{2}|Y}) \right)
\end{array}
\end{equation}
Hence
\begin{equation}
\begin{array}{lcl}
E&=&\displaystyle \iiint_{{\cal D}_{\bf Y}\times{\cal D}^2_{\bf X|Y}}  dy\,p_{\bf Y}(y)\;dx_{1}\,p_{\bf X|Y}(x_{1}|y)\; dx_{2}\, p_{\bf X|Y}(x_{2}|y) \;x_{1}x_{2}\\
&=&{\cal E}_{{\bf Y},{\bf (X_1,\,X_2)|Y}}\left( {\bf X_{1}X_{2}|Y}\right)
\end{array}
\end{equation}

The same demonstration can be made to establish that:
\begin{equation}
{\cal E}_{\bf Y}({\cal E}_{\bf X|Y}({\bf X|Y})^{n})={\cal E}_{{\bf Y},{\bf (X_1,\,X_2, ...,\,X_n)|Y}}({\bf X_{1}~X_{2} ...X_{n}|Y})
\end{equation}

Let us now assume that the weights associated with the random variable ${\bf Y}$ are described by a nonlinear function ${\bf f}({\bf Z_{\bf Y}})$ of the conditional expectation ${\bf Z_{\bf Y}} = {\cal E}_{\bf X|Y} \left( {\bf X|\bf Y}  \right) $. 
The problem then becomes to compute:
\begin{equation}
E={\cal E_{\bf Y}} {\bf \left( f(Z_{\bf Y}) \right) }= {\cal E}_{\bf Y} \left({\bf f}\left( {\cal E}_{\bf X|Y} \left( {\bf X|\bf Y}  \right) \right) \right)
\label{eq:form}
\end{equation}

Such a nonlinearity can be treated with no approximation using a projection on an infinite basis. 
In all the examples presented in this article, we have used a Taylor polynomials basis, which means that ${\bf f}(x)$ is expanded around $x_{0}$
\begin{equation}
{\bf f}(x)=\sum_{n=0}^{+\infty}\frac{\partial^{n} {\bf f}(x_{0})}{n!}\left(x -x_{0}\right)^n
\label{eq:taylor}
\end{equation}
We note that both $x_{0}$ and ${\bf f}$ can be conditioned by $\bf Y$.
Now, following the same line as explained above for the Basics of Monte Carlo Methods, we regard the sum in the expansion of ${\bf f}$ as an expectation, writing:
\begin{equation}
{\bf f}(x)=  {\cal E}_N \left( \frac{\partial^{N}{\bf f}(x_{0})}{P_N(N) N!}\left(x -x_{0}\right)^N \right)
\label{eq:taylor2}
\end{equation}
where the random variable $N$ (of probability law $P_{N}$) is the degree of one monomial in the Taylor polynomial.
This step only requires us to define one infinite set of probabilities (instead of two in Eq.~\ref{eq:somme}), with $\sum_{n=0}^{+\infty} P_N(n) = 1$. 

Equation~\ref{eq:form} can then be rewritten as:
\begin{equation}
E= {\cal E}_{\bf Y} \left( {\bf f}\left( {\cal E}_{\bf X|Y} \left( {\bf X|\bf Y}  \right) \right) \right)
= {\cal E}_{Y,N} \left( \frac{\partial^{N}{\bf f}(x_{0})}{P_N(N) N!}\left( {\cal E}_{\bf X|Y}({\bf X|Y})   - x_{0}\right)^N\right)
\label{eq:taylor3}
\end{equation}
Defining independent and identically distributed random variables ${\bf X}_q$, with the same distribution as ${\bf X}$, the innermost term rewrites
\begin{equation}
E={\cal E}_{Y,N} \left(\frac{\partial^{N}{\bf f}(x_{0})}{P_N(N) N!}\prod_{q=1}^{N}\left({\cal E}_{\bf X_q|Y}({\bf X}_{q}|{\bf Y}) -x_{0}\right)\right)
\label{eq:taylor4}
\end{equation}
or, equivalently:
\begin{equation}
E={\cal E}_{Y,N} \left(\frac{\partial^{N}{\bf f}(x_{0})}{P_N(N) N!}\prod_{q=1}^{N}{\cal E}_{\bf X_q|Y}({\bf X}_{q}|{\bf Y}-x_{0}) \right)
\label{eq:taylor4b}
\end{equation}
Since the variables ${\bf X}_{q}|{\bf Y}$ are independent in the innermost term, we have:
\begin{equation}
\prod_{q=1}^{N}{\cal E}_{\bf X_q|Y}({\bf X}_{q}|{\bf Y}-x_{0}) ={\cal E}_{\bf (X_1\,,X_2,...,X_N)|Y}\left( \prod_{q=1}^{N}\left({\bf X}_{q}|{\bf Y}-x_{0}\right) \right)
\label{eq:taylor4c}
\end{equation}
so that:
\begin{equation}
E={\cal E}_{\bf Y,N} \left(\frac{\partial^{N}{\bf f}(x_{0})}{P_N(N) N!}\,{\cal E}_{\bf (X_1\,,X_2,...,X_N)|Y}\left( \prod_{q=1}^{N}\left({\bf X}_{q}|{\bf Y}-x_{0}\right) \right)\right)
\label{eq:taylor5}
\end{equation}
and we finally have:
\begin{equation}
E={\cal E}_{\bf Y,N,(X_1\,,X_2,...,X_N)|Y} \left(\frac{\partial^{N}{\bf f}(x_{0})}{P_N(N) N!}\prod_{q=1}^{N}\left({\bf X}_{q}|{\bf Y}-x_{0}\right)\right)
\label{eq:taylor5a}
\end{equation}
which can be read as:
\begin{equation}
E={\cal E}_{\bf Y,N,(X_1\,,X_2,...,X_N)|Y} \left( w({\bf Y},{\bf N},{\bf X_{1}},{\bf X_{2}},...,{\bf X_{N}})\right)
\label{eq:taylor5b}
\end{equation}
with
\begin{equation}
w({\bf Y},{\bf N},{\bf X_{1}},{\bf X_{2}},...,{\bf X_{N}})=\frac{\partial^{N}{\bf f}(x_{0})}{P_N(N) N!}\prod_{q=1}^{N}\left({\bf X}_{q}|{\bf Y}-x_{0}\right)
\label{eq:taylor5c}
\end{equation}
With the notation above, $\prod_{q=1}^{0}\left({\bf X}_{q}|{\bf Y}-x_{0}\right)=1$.

The translation into a Monte Carlo algorithm then follows: 
\vspace{-0.8cm}
\begin{itemize}
\item sample a realisation ${\bf y}$ of ${\bf Y}$ (and set $x_{0}$ and ${\bf f}$ accordingly if they depend on ${\bf y}$)
\item sample a realisation $n$ of $N$
\item sample $n$ independent realisations ${\bf x}_{q=1,...,n}$ of ${\bf X}$ conditioned by $\bf y$
\item keep 
$$ \hat{w}=w({\bf y}, n, {\bf x}_{1}, ..., {\bf x}_{n}) = \frac{\partial^{n}{\bf f}(x_{0})}{P_{N}(n) n!}\prod_{q=1}^{n}\left({\bf x}_{q} -x_{0}\right) $$
\end{itemize}
and estimate $E$ as the average of many realisations $\hat{w}$.

\paragraph{Implementation example.} Let us illustrate the choice of the discrete distribution $P$ on $N$ with an implementation example. We take $Y$ uniformly distributed over $]0,1[$, $X|Y$ uniformly distributed over $]0,Y[$ and $f(x) = 1/(1+x)$ ($f$ corresponds to the photobioreactor real-world example in Figure 1.c, with $C=1$, $\alpha=0$, $\beta=-1$, $K_r=1$). Equation~\ref{eq:form} becomes
\begin{equation}
E = {\cal E}_{Y} \left(\dfrac{1}{1+{\cal E}_{X|Y}}\right)
\label{eq:implementationExample}
\end{equation}
Its analytical solution is $E = 2\ln\left(3/2\right)$.

Injecting the n-th derivative $\partial^{n}f(x_{0})=\dfrac{n!(-1)^n}{(1+x_0)^{n+1}}$ into Equation~\ref{eq:taylor5c} leads to
\[
w(Y,N,X_{1},X_{2},...,X_{N})=\frac{(-1)^N}{P_N(N) (1+x_0)^{N+1}}\prod_{q=1}^{N}\left(X_{q}|Y-x_{0}\right)
\]
that can be reformulated as
\begin{equation}
w(Y,N,X_{1},X_{2},...,X_{N})=\frac{1}{P_N(N)}\frac{x_0^N}{(1+x_0)^{N+1}}\prod_{q=1}^{N}\frac{X_{q}|Y-x_{0}}{x_0}
\label{eq:implemExempEstim1}
\end{equation}
Using standard importance-sampling reasoning, we choose the set of probabilities that cancels the term $\frac{x_0^N}{(1+x_0)^{N+1}}$ in the estimator:
\begin{equation}
w(Y,N,X_{1},X_{2},...,X_{N})=\prod_{q=1}^{N}\frac{X_{q}|Y-x_{0}}{x_0}
\label{eq:implemExempEstim1Simlified}
\end{equation}
with
\begin{equation}
P_N(N)=\frac{x_0^N}{(1+x_0)^{N+1}}
\label{eq:implemExempProba1}
\end{equation}

The NLMC algorithm is
\begin{itemize}
\item sample a realisation $y$ of $Y$
\item sample a realisation $n$ of $N$ according to the discrete distribution in Equation~\ref{eq:implemExempProba1}
\item sample $n$ independent realisations $x_{q=1,...,n}$ of $X$ uniformly distributed over $]0,y[$
\item keep 
$$ \hat{w}= \prod_{q=1}^{n}\frac{x_{q}-x_{0}}{x_0} $$
\end{itemize}
and estimate $E$ as the average of $M$ realisations $\hat{w}$. 

We define the computational cost $C$ of this algorithm in terms of the total number of random generations that are required to achieve $1\%$ standard deviation on the estimation. Each realisation of the algorithm includes 1 random generation $y$ of $Y$, 1 random generation $n$ of $N$, and $n$ random generations of $X$. And it takes $M_{1\%}$ realisations of the algorithm to achieve a standard deviation of $1\%$. Overall, 
\begin{equation}
C = M_{1\%}(2+{\cal E}(N)) = M_{1\%}(2+x_0)
\label{eq:implemExempCost}
\end{equation}
since ${\cal E}(N)=x_0$ with the discrete distribution in Equation~\ref{eq:implemExempProba1}. Figure 3 shows the values of $M_{1\%}$ and $C$ recorded with simulations, as a function of $x_0$. The choice of $x_0$ alone controls both the statistical convergence ({\it i.e.} $M_{1\%}$) and the computational cost (through the discrete distribution $P$ on $N$). We observe a trade-off between estimation and computational cost. For low values of $x_0$, only few realisations of $X$ are needed since the discrete distribution on $N$ is rapidly decreasing with $n$, but a large number of realisation of the algorithm are required for the estimation ({\it i.e.} ${\cal E}(N)$ is small but $M_{1\%}$ is large). At the opposite, for larger values of $x_0$, the estimator converges rapidly, but the average number of $X$ random generations per Monte Carlo realisations is increased ({\it i.e.} $M_{1\%}$ is small but ${\cal E}(N)$ is large). In between, we observe an optimal choice of $x_0$.

\clearpage
Figure 3 --- \textbf{Choice of the discrete distribution $P$ on $N$: a tradeoff between estimation and computational cost.}

\vspace{2cm}

\includegraphics[width=13cm]{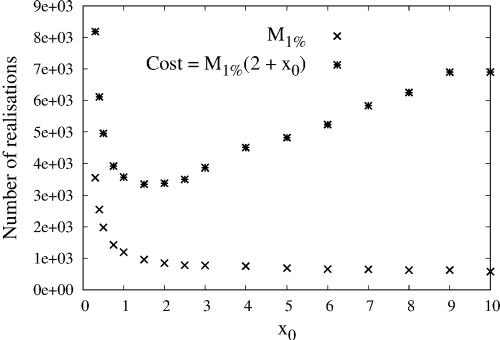}

\bigskip

\paragraph{Fig. 3 \textbf{Choice of the discrete distribution $P$ on $N$: a tradeoff between estimation and computational cost.}}
$\;$\\
Estimation of equation~\ref{eq:implementationExample} with the estimator in equation~\ref{eq:implemExempEstim1Simlified}, as a function of the choice of the fixed point $x_0$ around which the nonlinear function is Taylor expanded. In this example, $x_0$ also defines the discrete distribution $P_N$ (see equation~\ref{eq:implemExempProba1}). Crosses indicate the number of realisations $M_{1\%}$ that are required to achieve an estimation with 1\% standard deviation. Stars indicate the corresponding computational cost as defined in equation~\ref{eq:implemExempCost}.
\clearpage

\paragraph{Comparison with the naive plug-in estimator and convergence issues.} In the previous implementation example solving equation~\ref{eq:implementationExample}, a naive plug-in estimator could be constructed as\cite{NLexpectation}:
\[
E = {\cal E}_{Y} \left(\dfrac{1}{1+\dfrac{1}{K} \sum_{q=1}^{K} X_{q}|Y}\right)
\]
leading to the following Monte Carlo algorithm:
\begin{itemize}
\item sample a realisation $y$ of $Y$
\item sample $K$ independent realisations $x_{q=1,...,K}$ of $X$ uniformly distributed over $]0,y[$
\item keep 
$$ \hat{w}= \dfrac{1}{1+\dfrac{1}{K} \sum_{q=1}^{K} x_{q}} $$
\end{itemize}
and estimate $E$ as the average of $M$ realisations $\hat{w}$. With this algorithm, $K$ must ensure that the bias of the estimator can be neglected. For that purpose, we choose the value $K_{1\%}$ that always gives an estimation of ${\cal E}_{X|Y}$ with 1\% standard deviation. Therefore, each realisation of the algorithm includes 1 random generation for $Y$ and $K_{1\%}$ generations for $X$, and it takes $M_{1\%}$ realisations of the algorithm to achieve a standard deviation of $1\%$ on the estimation of $E$. The computational cost of the naive algorithm is therefore $M_{1\%}(1+K_{1\%})$. In the present example, $K_{1\%}=3333$ and we observed that $M_{1\%} = 140$ thanks to numerical simulations: the cost is $C = 466760$. Compared to the results in Figure 3, even for this very simple example where $Y$ and $X|Y$ have little variance, the computational cost of the naive plug-in algorithm is 100 times higher than that of the NLMC estimator. Nevertheless, this conclusion only stands for a reasonable choice of $x_0$ (and therefore of $P_N$). Indeed, the computational cost with the NLMC estimator seems to rise up to infinity when $x_0$ approaches 0 (see Equation~\ref{eq:implemExempEstim1Simlified} and Figure 3): even the naive plug-in algorithm would then be a better choice. Although we did not further analyse this observation with theoretical means, we can at least retain that chosing $x_0$ is likely to be an essential step of the present approach as far as computational costs are concerned. 

\clearpage

\section{Supplemental Information}

\clearpage

\renewcommand{\v}[1]{\ensuremath{\mathbf{#1}}} 
\newcommand{\gv}[1]{\ensuremath{\mbox{\boldmath$ #1 $}}} 
\newcommand{\parent}[1]{\left( #1 \right)} 
\newcommand{\accol}[1]{\left\lbrace #1 \right\rbrace} 
\newcommand{\croch}[1]{\left[ #1 \right]} 


\section{{SI1} --- Solar thermochemical reduction of zinc oxide averaged over the year}

\subsection{The problem} 
Solar-driven high temperature thermochemical cycles processes, commonly based on metal oxides reduction~\cite{SPST05A,SPMU08A}, are an alternative to fossil fuel-based method for $H_2$ generation. 
Here we focus on thermal reduction zinc oxide, as the first part of a two step water splitting cycle.
Photons emitted from the sun are reflected on heliostats and  concentrated at the entrance of the chemical reactor in which $\mathrm{ZnO}$ dissociation is carried out. 
Solar power received by the reactor at a given instant determines the chemical conversion rate of the reaction $\mathrm{ZnO \rightarrow Zn + \frac{1}{2} O}$. 
The present Monte Carlo estimation of the solar-plant's annual conversion rate $C$ associates the thermochemical knowledge of the non-linear kinetics of zinc oxide dissociation~\cite{SPSC09A,SHPI11A} to the description of radiative transfer on a multiple-reflection solar receiver~\cite{LSDE13A}.
The nonlinearity lies in the instantaneous coupling between photon transport and zinc-oxide reduction.

The random variable ${\bf X}$ is the contribution of an optical path from the sun to the entrance of the chemical reactor. 
Its expectation ${\cal E}_{{\bf X}|{\bf Y}}( {\bf X}|{\bf Y} )$ is the instantaneous thermal-power fraction collected at a given moment $\bf Y$ of the year\cite{Farges15}.
Solar power is used to reduce the zinc oxide with a nonlinear conversion rate $\mathrm{f}\,({\cal E}_{{\bf X}|{\bf Y}}( {\bf X}|{\bf Y} ))$ (see\cite{SPSC09A,SHPI11A}).

\subsection{Non-Linear Monte Carlo formulation} 
The annual conversion rate $C$ is reformulated as its Taylor expansion around $x_0$.
First, independent and identically distributed (i.i.d) optical-path random variables are introduced, each defining a random contribution ${\bf X}_i|{\bf Y}$ i.i.d as ${\bf X}|{\bf Y}$. Then, we expand the non-linear function $\mathrm{f}$ around $x_0$ chosen as an upper bound of ${\bf X}|{\bf Y}$. Finally, the infinite power series is statistically formulated thanks to the discrete random variable $N$ the order of Taylor expansion:
\begin{equation}
N_{\bf Y} = B_{{\bf X}_1|{\bf Y}}+\sum_{i=2}^{+\infty}\,i\,B_{{\bf X}_i|{\bf Y}}\,\prod_{q=1}^{i-1}(1-B_{{\bf X}_q|{\bf Y}})
\end{equation}
where the $B_{{\bf X}_i|{\bf Y}}$ are Bernoulli random variables:
\begin{equation}
\begin{array}{ll}
B_{{\bf X}_i|{\bf Y}} = \left\lbrace
\begin{array}{l}
\displaystyle 1 \hspace{1cm} \mbox{with probability} \hspace{4mm} P_{{\bf X}_i|{\bf Y}}=\frac{{\bf X}_i|{\bf Y}}{x_0} \vspace{3mm}\\
\displaystyle 0 \hspace{1cm} \mbox{with probability} \hspace{4mm} 1-P_{{\bf X}_i|{\bf Y}}
\end{array}\right.
\end{array}
\label{eq:SIPSolBernou}
\end{equation}

In the end, the reformulation is:
\begin{equation}
C = {\cal E}_{{\bf Y},({\bf X}_1,{\bf X}_2,...,{\bf X}_{{\bf N}_{\bf Y}})|{\bf Y}}\left(w(N_{\bf Y})\right)
\end{equation}
where the Monte Carlo weight function $w$ is
\begin{equation}
w(u)=\left\lbrace\begin{array}{l}
\displaystyle \exp\left(C_{4}\,x_0^{-\alpha}\right) \displaystyle\sum_{m=1}^{3}C_{m}\,x_0^{(m\, \alpha-1)} \;\;\mbox{if}\; u=1\vspace{6mm} \\
w(1)+ \exp\left(C_{4}\,x_0^{-\alpha}\right)\displaystyle\sum_{k=0}^{u}(-1)^k \displaystyle\sum_{p=0}^{k}\frac{1}{p!(k-p)!}\displaystyle\sum_{m=1}^{3}C_{m}\,x_0^{(m \alpha-1+p)}\\
\hspace{3cm}\times\displaystyle\prod_{j=1}^{k-p}(m \alpha-j)\displaystyle\sum_{r=1}^{p}(-1)^r\,C_4\,x_0^{-(\alpha+r)}\prod_{l=0}^{r-1}(\alpha+l) \;\;\mbox{if}\; u>1
\end{array}\right.
\label{eq:SISOLw}
\end{equation}
with $u\in\mathbb{N}$ and $C_1$, $C_2$, $C_3$, $C_4$, $\alpha$ and $x_0$ known constants.

\subsection{Algorithm}

\begin{description}
\item[Step 1] Uniform sampling of a moment ${\bf y}$ of the year and index initialisation $n=1$.
\item[Step 2] Sampling of the $n$-th optical path contributing to thermal-power collection at instant ${\bf y}$ and computation of its contribution $x_n$ (as presented in\cite{LSDE13A,Farges15}).
\item[Step 3] Computation of the probability $P_n$ in Eq.~\ref{eq:SIPSolBernou} and sampling of a realisation $b_n$ of the Bernoulli random variable $B_{{\bf X}_n|{\bf Y}}$.
\item[Step 4] If $b_n=0$, the algorithm loops to step 2 after incrementation of $n$. Else, the procedure is terminated and the weight ${\hat w}(n)$ computed according to Eq.~\ref{eq:SISOLw}.
\end{description}

\subsection{Simulated configuration}
Results shown in Extended Data Figure 1 are obtained for a 1 MW solar plant with a 80 m high central receiver tour and 160 heliostats arranged in a radial stagered layout (nueen method). Values of the conversion-rate parameters $C_1$, $C_2$, $C_3$, $C_4$ and $\alpha$ are given in~\cite{SPSC09A,SHPI11A}. Meteonorm DNI database.

\clearpage

\section{{SI2}  --- Wave scattering}

This example is fully detailed in Charon {\it et al.} 2016\cite{Charon_16}.

\subsection{The problem}
Here we address the solution of Schiff's approximation\cite{Schiff}, also known as the {\it anomalous diffraction approximation}\cite{VdH}, for an incident plane wave with propagation direction ${\bf e}_i$ and wave number $k$ scattered by the scattering potential $U$ that takes value $U^{in}$ inside a large domain compared to $1/k$ and $0$ outside (this domain corresponds to the shape of the scatterer), with $\vert U^{in} \vert \ll k^2$. Let $\Gamma_{\bf Y}$ be the projected surface of the scatterer seen from ${\bf e}_i$, given the scatterer orientation ${\bf Y}$. Straight rays along ${\bf e}_i$ cross $\Gamma_{\bf Y}$ at locations ${\bf P}|{\bf Y}$; the domain of definition of ${\bf P}|{\bf Y}$ is then ${\cal D}_{{\bf P}|{\bf Y}}\equiv {\bf P}|{\bf Y}$. These rays are attenuated and phase shifted over the crossing length $l({\bf P}|{\bf Y})$. The complex random variable $X|{\bf Y}=X_{{\bf P}|{\bf Y}}({\bf e}_s)$ is the contribution of one of these rays to the scattered field in a direction ${\bf e}_s$ deviating from ${\bf e}_i$ by a small scattering angles $\theta$\cite{Charon_16}:
\begin{equation}
X_{{\bf P}|{\bf Y}}({\bf e}_s) = A_{\bf Y}\frac{k}{2\,\pi}\exp(ik\,\theta\,{\bf b}\cdot{\bf P}|{\bf Y})\left\lbrace 1 - \exp\left(\frac{i}{2k}U^{in}\,l({\bf P}|{\bf Y})\right) \right\rbrace
\label{eq:SIWaveX}
\end{equation}
where ${\bf b}$ is the unit vector along the projection of ${\bf e_s}$ on the plane containing ${\cal D}_{{\bf P}|{\bf Y}}$ and $A_{\bf Y}$ is the area of ${\cal D}_{{\bf P}|{\bf Y}}$. Given the orientation ${\bf Y}$ of the scatterer, the conditional expectation $S_{\bf Y}({\bf e}_s)={\cal E}_{X|{\bf Y}}(X|{\bf Y})={\cal E}_{{\bf P}|{\bf Y}}(X_{{\bf P}|{\bf Y}}({\bf e}_s))$ is the complex scattering amplitude in direction ${\bf e}_s$\cite{Schiff,VdH,Dauchet_14,Charon_16}. For the scatterer orientation ${\bf Y}$, the far-field scattering diagram is given by the \emph{differential scattering cross-section}\cite{VdH,Bohren}: ${\hat W}_{\bf Y}({\bf e}_s)=\vert S_{\bf Y}({\bf e}_s) \vert^2=\vert {\cal E}_{{\bf P}|{\bf Y}}(X_{{\bf P}|{\bf Y}}({\bf e}_s)) \vert^2$.

We address the Monte Carlo computation of $W({\bf e}_s) = {\cal E}_{\bf Y}({\hat W}_{\bf Y}({\bf e}_s))$ which is ${\hat W}_{\bf Y}({\bf e}_s)$ averaged over the statistics of the scatterer orientation ${\bf Y}$~\cite{VdH,Bohren,Dauchet_14}. 
The full expression is then
\begin{equation}
\begin{array}{lcl}
W({\bf e}_s) & = & {\cal E}_{\bf Y}\left( \vert S_{\bf Y}({\bf e}_s) \vert^2 \right)\\
& = &{\cal E}_{\bf Y} \left( \vert {\cal E}_{{\bf P}|{\bf Y}}(X_{{\bf P}|{\bf Y}}({\bf e}_s)) \vert^2 \right) \\
& = &{\cal E}_{\bf Y} \left( \left(\Re \, {\cal E}_{{\bf P}|{\bf Y}}(X_{{\bf P}|{\bf Y}}({\bf e}_s)) \right)^2 + \left(\Im \, {\cal E}_{{\bf P}|{\bf Y}}(X_{{\bf P}|{\bf Y}}({\bf e}_s))\right)^2 \right)
\end{array}
\end{equation}
where the configuration spaces are the orientation vectors of the scattering potential with respect to ${\bf e}_i$ (${\cal D}_{\bf Y}$) and the projected surface of the scattering potential seen from the incident direction ${\bf e}_i$ (${\cal D}_{{\bf P}|{\bf Y}}$).

\subsection{Non-Linear Monte Carlo formulation} 

$W({\bf e}_s)$ is reformulated based on the definition of two independent and identically distributed location random-variables ${\bf P}_1|{\bf Y}$ and ${\bf P}_2|{\bf Y}$:
\begin{equation}
\begin{array}{lcl}
W({\bf e}_s) &=&{\cal E}_{\bf Y} \left( \Re \, {\cal E}_{{\bf P}_1|{\bf Y}}(X_{{\bf P}_1|{\bf Y}}({\bf e}_s)) \; \Re \, {\cal E}_{{\bf P}_2|{\bf Y}}(X_{{\bf P}_2|{\bf Y}}({\bf e}_s))   + \Im \, {\cal E}_{{\bf P}_1|{\bf Y}}(X_{{\bf P}_1|{\bf Y}}({\bf e}_s)) \; \Im \, {\cal E}_{{\bf P}_2|{\bf Y}}(X_{{\bf P}_2|{\bf Y}}({\bf e}_s)) \right)\\
&=& {\cal E}_{{\bf Y},({\bf P}_1,{\bf P}_2)|{\bf Y}}\left(w({\bf P}_1,{\bf P}_2,{\bf e}_s) \right)
\end{array}
\end{equation}
where the Monte Carlo weight function $w$ is:
\begin{equation}
w({\bf P}_1,{\bf P}_2,{\bf e}_s) = \Re\,X_{{\bf P}_1|{\bf Y}}({\bf e}_s)\;\Re\,X_{{\bf P}_2|{\bf Y}}({\bf e}_s) + \Im\,X_{{\bf P}_1|{\bf Y}}({\bf e}_s)\;\Im\,X_{{\bf P}_2|{\bf Y}}({\bf e}_s)
\label{eq:SIWaveW}
\end{equation}
with $X_{{\bf P}_q|{\bf Y}}({\bf e}_s)$ defined in Eq.~\ref{eq:SIWaveX}.

\subsection{Algorithm}
The sampling procedures of the Monte-Carlo algorithm are then:
\begin{description}
\item[Step 1] Isotropic sampling of an orientation ${\bf y}$ of the scattering potential.

\item[Step 2] Uniform sampling of the first location ${\bf p}_1$ on the projected surface defined by ${\bf y}$ and computation of the corresponding crossing length $l({\bf p}_1)$: the realisation $x_1$ of $X_{{\bf P}|{\bf Y}}({\bf e}_s)$ is computed according to Eq.~\ref{eq:SIWaveX}.

\item[Step 3] Uniform sampling of the second location ${\bf p}_2$ on the projected surface defined by ${\bf y}$ and computation of the corresponding crossing length $l({\bf p}_2)$: the realisation $x_2$ of $X_{{\bf P}|{\bf Y}}({\bf e}_s)$ is computed according to Eq.~\ref{eq:SIWaveX}.

\item[Step 4] Computation of the weight ${\hat w}=w({\bf p}_1,{\bf p}_2,{\bf e}_s)$ according to Eq.~\ref{eq:SIWaveW}: ${\hat w}=\Re\,x_1\;\Re\,x_2 + \Im\,x_1\;\Im\,x_2$
\end{description}

Codes for the implementation of this algorithm in the case of spheroidal and cylindrical scattering potentials, as well as validation against reference solution, are provided in Charon et al. 2016 \cite{Charon_16} (see also http://edstar.lmd.jussieu.fr/codes). 

\subsection{Simulated configuration}
Results shown in Extended Data Figure 2 are obtained for $k=2\pi/\lambda$, with wavelength $\lambda=500 nm$ and the scattering potential $U^{in} = -2k(m-1)$, with $m = 1.2 -i\,4\,.10^{-3}$ (scatterer relative refractive index\cite{Dauchet_14}). The shape of the scatterer is helical with length $L=50 \mu m$, pitch $P = 15 \mu m$, helix diameter $D=20\mu m$, cylinder diameter $d=3.5\mu m$.

\clearpage
\section{{SI3}  --- Phytoplankton growth in light-limited environments}

\subsection{The problem} 
Here we address the production of reference solutions for a photobioreactor model. This model is based on a radiative transfer approach presented in\cite{Dauchet2016}. The rate of photon absorption ${\bf A}_{\bf Y}$ by micro-algae at location ${\bf Y}$ within the culture volume is solution of the Radiative Transfer Equation. The present study is based on the standard linear transport MC algorithm presented in\cite{Dauchet_13} for the estimation of ${\bf A}_{\bf Y}$ at any location within the photobioreactor. In this algorithm, realisations of the optical path random variable ${\bf \Gamma}(C)|{\bf Y}$ are sampled backward from the absorption location ${\bf Y}$ to the light emitting surface (the surface of the 979 light-diffusing optical fibres, see EDF3). Since micro-algae are scattering visible light, ${\bf \Gamma}(C)|{\bf Y}$ depends on their concentration $C$ within the suspension: the scattering and absorption coefficient of the suspension are proportional to $C$ (independent scattering). The random variable ${\bf X}(C)|{\bf Y}={\bf X}_{{\bf \Gamma}(C)|{\bf Y}}(C)$ is the contribution of one of these optical paths to the photon absorption rate at ${\bf Y}$ (see\cite{Dauchet_13} for the detailed expression of ${\bf X}_{{\bf \Gamma}(C)|{\bf Y}}(C)$). This contribution ${\bf X}_{{\bf \Gamma}(C)|{\bf Y}}(C)$ depends on the biomass concentration $C$ because the absorption coefficient of the suspension determines the fraction of incident light flux that is transmitted along the optical path ${\bf \Gamma}(C)|{\bf Y}$ (according to Beer law). In the end, the conditional expectation ${\bf A}_{\bf Y}(C)={\cal E}_{{\bf X}|{\bf Y}}({\bf X}|{\bf Y})={\cal E}_{{\bf \Gamma}(C)|{\bf Y}}({\bf X}_{{\bf \Gamma}(C)|{\bf Y}}(C))$ is the rate of photon absorption at location ${\bf Y}$, for the biomass concentration $C$.

Photons absorbed by a micro-algae are converted within the photosynthetic units and the Z-scheme\cite{Bio,Cornet_09}, leading to the local biomass growth-rate\cite{Takache_12, Dauchet2016}
\begin{equation}
\begin{array}{lcl}
r_{\bf Y}(C) &=& \mathrm{f}({\bf A}_{\bf Y}(C),C)\\
&=& C \left({\alpha\;\frac{K\,{\bf A}_{\bf Y}(C)}{K+{\bf A}_{\bf Y}(C)} - \beta\;\frac{K_r}{K_r+{\bf A}_{\bf Y}(C)}} \right)\\
&=&C \left({\alpha\;\frac{K\,{\cal E}_{{\bf \Gamma}(C)|{\bf Y}}({\bf X}_{{\bf \Gamma}(C)|{\bf Y}}(C))}{K+{\cal E}_{{\bf \Gamma}(C)|{\bf Y}}({\bf X}_{{\bf \Gamma}(C)|{\bf Y}}(C))} - \beta\;\frac{K_r}{K_r+{\cal E}_{{\bf \Gamma}(C)|{\bf Y}}({\bf X}_{{\bf \Gamma}(C)|{\bf Y}}(C))}} \right)
\end{array}
\label{eq:SIPBRLoi}
\end{equation}
where $\alpha$, $\beta$, $K$ and $K_r$ are constant parameters that depend on the studied microorganism (see\cite{Takache_12} for values in the case of {\it Chlamydomonas Reinhardtii}).

Due to light absorption and scattering by micro-algae, the field ${\bf A}_{\bf Y}$ is heterogeneous within the volume (less light farther from the fibres and for higher concentrations) and so is the local photosynthetic rate $r_{\bf Y}(C)$. We address the Monte Carlo computation of $R(C)={\cal E}_{\bf Y}(r_{\bf Y}(C))$, the local photosynthetic rate $r_{\bf Y}(C)$ averaged over locations ${\bf Y}$ of the culture volume. The full expression is then
\begin{equation}
R(C)={\cal E}_{\bf Y}\left( C \left({\alpha\;\frac{K\,{\cal E}_{{\bf \Gamma}(C)|{\bf Y}}({\bf X}_{{\bf \Gamma}(C)|{\bf Y}}(C))}{K+{\cal E}_{{\bf \Gamma}(C)|{\bf Y}}({\bf X}_{{\bf \Gamma}(C)|{\bf Y}}(C))} - \beta\;\frac{K_r}{K_r+{\cal E}_{{\bf \Gamma}(C)|{\bf Y}}({\bf X}_{{\bf \Gamma}(C)|{\bf Y}}(C))}} \right) \right)
\label{eq:SIPBRFullExpr}
\end{equation}

\subsection{Non-Linear Monte Carlo formulation} 

$R(C)$ is reformulated based on the Taylor expansion of $\mathrm{f}$ around $x_0$. First, independent and equally distributed optical-path random variables ${\bf \Gamma}_i(C)|{\bf Y}$ are introduced, defining independent and equally distributed contributions ${\bf X}_i(C)|{\bf Y}={\bf X}_{{\bf \Gamma}_i(C)|{\bf Y}}(C)$. Then, we expand the non-linear function in Eq.~\ref{eq:SIPBRLoi} around  $x_0$ chosen as an upper bound of ${\bf X}_i(C)|{\bf Y}$. Finally, the infinite power series is statistically formulated thanks to the discrete random variable $N$ the order of Taylor expansion:
\begin{equation}
N_{\bf Y}(C) = B_{{\bf X}_1(C)|{\bf Y}}+\sum_{i=2}^{+\infty}\,i\,B_{{\bf X}_i(C)|{\bf Y}}\,\prod_{q=1}^{i-1}(1-B_{{\bf X}_q(C)|{\bf Y}})
\end{equation}
where the $B_{{\bf X}_i(C)|{\bf Y}}$ are Bernoulli random variables:
\begin{equation}
\begin{array}{ll}
B_{{\bf X}_i(C)|{\bf Y}} = \left\lbrace
\begin{array}{l}
\displaystyle 1 \hspace{1cm} \mbox{with probability} \hspace{4mm} P_{{\bf X}_i(C)|{\bf Y}}=\frac{{\bf X}_i(C)|{\bf Y}}{x_0} \vspace{3mm}\\
\displaystyle 0 \hspace{1cm} \mbox{with probability} \hspace{4mm} 1-P_{{\bf X}_i(C)|{\bf Y}}
\end{array}\right.
\end{array}
\label{eq:SIPBRBernou}
\end{equation}

In the end, the reformulation is:
\begin{equation}
R(C) = {\cal E}_{{\bf Y},({\bf \Gamma}_1,{\bf \Gamma}_2,...,{\bf \Gamma}_{{\bf N}_{\bf Y}})|{\bf Y}}\left(w(N_{\bf Y})\right)
\end{equation}
where the Monte Carlo weight function $w$ is
\begin{equation}
w(u)=\left\lbrace\begin{array}{l}
\displaystyle C\left[\alpha\;\frac{K\,x_0}{K+x_0} - \beta\;\frac{K_r}{K_r+x_0}\right] \;\;\mbox{if}\; u=1\vspace{6mm} \\
w(1)-\displaystyle\sum_{i=1}^{u-1}C\left[\alpha\;\frac{K^2}{K+x_0}\left(\frac{x_0}{K+x_0}\right)^i + \beta\;\frac{K_r}{K_r+x_0}\left(\frac{x_0}{K+x_0}\right)^i\;\right] \;\;\mbox{if}\; u>1
\end{array}\right.
\label{eq:SIPBRpoids}
\end{equation}
with $u\in\mathbb{N}$ and $\alpha$, $K$, $\beta$, $K_r$, $x_0$ and $C$ known constants.

\subsection{Algorithm}

\begin{description}
\item[Step 1] Uniform sampling of a location ${\bf y}$ within the culture volume and index initialisation $n=1$.

\item[Step 2] Sampling of the $n$-th optical path contributing to absorption at ${\bf y}$: the realisation $x_n$ of ${\bf X}_{{\bf \Gamma}_n(C)|{\bf Y}}(C)$ is computed according to\cite{Dauchet_13}.

\item[Step 3] Computation of the probability in Eq.~\ref{eq:SIPBRBernou}, $P=x_n/x_0$, and sampling of a realisation $b_n$ of the Bernoulli random variable $B_{{\bf X}_n(C)|{\bf Y}}$.

\item[Step 4] If $b_n=0$, the algorithm loops to step 2 after incrementation of $n$. Else, the procedure is terminated and the weight ${\bf w}(n)$ computed according to Eq.~\ref{eq:SIPBRpoids}.
\end{description}

\subsection{Simulated configuration}
Results shown in Extended Data Figure 3 are obtained for kinetics parameters $\alpha = 1.785\,.10^{-9}$ $kg/\mu mol_{h\nu}$, $\beta = 4.057\,.10^{-6}$ $s^{-1}$, $K=32000$ $\mu mol_{h\nu}/kg/s$ and $K_r=7500$ $\mu mol_{h\nu}/kg/s$. The photobioreactor is a 25L DiCoFluV\cite{Cornet_10}: reactor diameter $16.5\,cm$, reactor height $1\,m$, optical fibres diameter $1.2\,mm$, fibres height $1\,m$, $979$ fibres, hexagonal lattice fibre arrangement with distance $4.8\,mm$ between two fibres axis (see\cite{Dauchet_13,Cornet_10}), homogeneous surface flux density $25\,\mu mol_{h\nu}/m^2/s$ emitted at fibres surface. The radiative properties (absorption and scattering by micro-algae) are those presented in\cite{Dauchet_14} for {\it Chlamydomonas Reinhardtii}.

\clearpage
\section{{SI4}  ---  Atmospheric radiative transfer: top-of-atmosphere specific intensity (from earth toward the outer space)}

This example is fully detailed in Galtier {\it et al.} 2015\cite{Galtier15}.

\subsection{The problem} 
Photons are emitted either by a surface (here the ground) or by the volume (the atmosphere). The radiation $I(\Delta\nu)$ perceived at the observation location TOA is the fraction of all these photons that have a frequency inside the observation-band $\Delta\nu$ and reach the sensor, i.e. which are not absorbed by the atmosphere in between. When no scattering occurs (clear sky), photon-paths are straight lines and the fraction of the photons of frequency $\nu_y$ that travel (let say along the strict vertical) from altitude $H_y$ to TOA is
$$exp(-\tau_y)$$
where
$$\tau_y=\int_{H_y}^{TOA} dx \sum_{j_x=1}^{N_t} h_{\nu_y,j_x}(x)$$
This exponential extinction is that of Beer law where $\tau_y$ is the monochromatic optical thickness, i.e. the sum of all transition cross-sections $h_{\nu_y,j_x}(x)$ at all intermediate altitudes $x$. In this example, the random variable $\bf Y$ is a vector that gathers the altitude $H_{\bf Y}$ of emission, the photon-path $\bf \Gamma_Y$ from $H_{\bf Y}$ to TOA and frequency $\nu_{\bf Y}$ of emission. Similarly, $\bf X$ gathers all the description of an absorption event, altitude $H_{\bf X}$ and index of absorption-transition $j_{\bf X}$.

In standard Monte Carlo approaches, $\tau_y$ is either precomputed and tabulated, or is easily computable from tabulated values of the absorption coefficient ($k = \sum_{j_x=1}^{N_t} h_{\nu_y,j_x}$). Then the Monte Carlo algorithm deals only with the sampling of $\bf Y$ and the weight function has the form
$$w(y)={\cal S}_y exp(-\tau_y)$$
${\cal S}_{\bf Y}$ is the source associated with the emission $\bf Y$ (Planck's function at the local temperature times an emissive power that depends on concentration, pressure and temperature). The sum over the $N_t$ transitions $j_x$ is thus not handled by standard Monte Carlo algorithms, despite the fact that this sum has all the features inviting to make use of statistical approaches:  $N_t$ is huge, typically of the order of $10^6$, and the deterministic pre-calculations can be computationally very demanding (and are to be re-performed when testing each new spectroscopic assumption). 
No attempt has been made so far to address this sum statistically, together with the photon-path statistics because these two statistics are combined via the non-linearity of the exponential extinction.

\subsection{Non-Linear Monte Carlo formulation} 

So the present question is to design a Monte Carlo algorithm performing the summation $\sum_{j=1}^{N_t} h_{\nu,j}$ together with the integrals over frequency, emission location and absorption location, despite of the nonlinearity of Beer extinction. This starts by looking at $\tau_y$ in statistical terms~: the integral over altitudes $H_{\bf X}$ is transformed into the sampling of absorption locations, and the sum $\sum_{j_x=1}^{N_t} h_{\nu_y,j_x}$ into the sampling of absorption-transitions. This double sampling is formally summarised into the random variable $\bf X$. To get a sample $x$ of $\bf X$, one first samples $H_{\bf X}$ and $j_{\bf X}$. Then $x$ is the value that the optical thickness $\tau_y$ would have if the atmosphere was homogeneous at the thermodynamic conditions of altitude $H_x$ and if all transitions were identical to $j_x$. This random variable depends on $\bf Y$ and its expectation, knowing ${\bf Y}=y$, is ${\cal E}({\bf X}| {\bf Y}=y) = \tau_y$. We therefore get $I\left(\Delta\nu \right) = {\cal E}_{\bf Y} \left( {\cal S}_{\bf Y}\; exp\right(-{\cal E}_{{\bf X}|{\bf Y}}({\bf X}|{\bf Y})\left) \right)$ 

\subsection{Algorithm}
A strict application of the solution described in {\it Methods} implies to first sample $\bf Y$ (as in a standard Monte Carlo algorithm), then sample the degree $n$ of one monomial in the Taylor expansion of the negative-exponential function $\mathrm{f}$, and draw $n$ independent samples $x_1, x_2 ... x_n$ of $\bf X$, i.e $n$ absorption locations $H_{x_q}$ (between emission location and TOA) and $n$ transition indexes $j_{x_q}$:
\begin{description}

\item[Step 1] sample a frequency,
\item[Step 2] sample an emission-altitude,
\item[Step 3] sample an order $n$ of the development of the exponential,
\item[Step 4] sample $n$ paired absorption-altitudes, transitions.
\end{description}

\subsection{Null-collision reformulation}
Instead of implementing this solution in this straightforward manner, we chose to use null-collisions\cite{Galtier13}. The same quantities are sampled, but the order is different, leading to quite intuitive physical pictures. The first step is still to sample a frequency, but then we implement a backward tracking algorithm, starting from the observation altitude. We sample a first collision-altitude as if photons were emitted at the observation location in the downward direction in an homogeneous atmosphere. Then we sample a transition and a statistical test is made to determine whether the algorithm stops at this altitude or continues (a Bernoulli test to determine whether the collision is a true one or a null-collision). If it continues, this first collision-altitude and first transition have the status of one of the $n$ absorption-altitudes and transitions in the above presented algorithm. Then a next collision-location is sampled, together with a next transition, etc. When the algorithm stops after $n$ collisions, the final altitude is interpreted as the emission-altitude. So $n$ and the emission-altitude are not sampled first: they are sampled by the successive Bernoulli tests as in a standard backward-tracking multiple-scattering algorithm.

\subsection{The final algorithm}
\begin{description}

\item[Step 1] Initialisation of the current altitude at $H_x\leftarrow TOA$.
\item[Step 2] Uniform sampling of a frequency $\nu_y$ in the considered infrared band (or the whole infrared).
\item[Step 3] Exponential sampling of a travelled distance $d$ before absorption for a photon of frequency $\nu_y$ travelling from $H_x$ in the backward direction within a virtual homogeneous atmosphere of absorption coefficient $\hat{k}_{\nu_y}$.
\item[Step 4] If the travelled distance leads the photon to hit the surface, $H_x\leftarrow 0$ and the algorithm jumps to {\it Step 7}. Otherwise $H_x\leftarrow H_x-d$.
\item[Step 5] Sampling of a state-transition $j_x$ according to ${\cal P}(1), {\cal P}(2) ... {\cal P}(N_t)$ (see\cite{Galtier15}).
\item[Step 6] Bernoulli trial of probability $P_a=\frac{h_{\nu_y,j_x}}{\hat{k}_{\nu_y} {\cal P}(j_x)}$ to decide whether the algorithm jumps to {\it Step 7} or {\it Step 3}.
\item[Step 7] $H_y\leftarrow H_x$ and the algorithm stops with $\hat{w}=\Delta\nu B_y$ where $B_y$ is the Planck function at frequency $\nu_y$ for the atmospheric temperature at $H_y$.
\end{description}

\subsection{Simulated configuration}
The simulation results in Extended Data Fig. 4 have been obtained using the HITRAN spectroscopic database. All other parameters are given in Galtier {\it et al.} 2015\cite{Galtier15}.

\subsection{Computational performance}
As far as computational costs are concerned, our computations did not require to first scan HITRAN and pre-compute absorption coefficients at all altitudes and all frequencies before running the Monte Carlo code for transfer since our Monte Carlo code handles both simultaneously despite the nonlinearity of the exponential extinction. The computational benefit is very significant when studying the effects of new spectroscopic data or new line-shape assumptions. Otherwise, for earth applications and fixed spectroscopic assumptions, there is no problem associated to the pre-computation of absorption coefficients and to their tabulation as function of molecular composition and thermodynamic state. But for combustion or astrophysics applications, thinking in particular of exoplanets, the diversity of compositions is extremely wide, temperatures can by high, imply the use of much larger spectroscopic databases, including hot lines, and pre-computation/tabulation of absorption coefficients is a today's challenge in itself. Our nonlinear Monte Carlo suppresses this need\cite{Galtier15}.

\clearpage
\section{{SI5} --- Gas kinetics}


\subsection{The problem}

We consider a gas of interacting particles, with collisions following Maxwell model, and a cross section $\sigma(\vec{c},\vec{c^*}) = \frac{\kappa}{4\pi \| \vec{c} - \vec{c^*} \|}$.
Particles are confined within an harmonic static trap of pulsation $\omega$ so that acceleration at position $\vec{r}$ is spring-like: $\vec{a}(\vec{r}) = -\omega^2 \vec{r}$.

We follow the distribution function $f(\vec{r},\vec{c},t)$ at location $\vec{r}$, velocity $\vec{c}$ at time $t$.
We consider it known at some time $t_I$, either constrained at local equilibrium, according to:

\begin{equation}
f_{_{\bf LEQ}} / n_I = p_{{\cal N}(\vec{u}_I, c_{q,I}^2)}
\end{equation}

or, out of equilibrium, according to:

\begin{equation}
f_{_{\bf BKW}} / n_I = \frac{1}{3} \frac{(\vec{c}-\vec{u}_I)^2}{\frac{c_{q,I}^2}{5/3}} \ p_{{\cal N}(\vec{u}_I, \frac{c_{q,I}^2}{5/3})}
\end{equation}

where $n_I$ denotes density,  $\vec{u}_I$ denotes the mean velocity and $c_{q,I}$ denotes the mean square speed at $\vec{r}$ at time $t_I$ ; $p_{{\cal N}(\mu,\sigma^2)}$ denotes probability density function of a Gaussian random variable, with expectation $\mu$ and variance $\sigma^{2}$.

\paragraph{For the case in fig. 3a,} we set $\omega = 0$, and $n_I$, $\vec{u}_I$ and $c_{q,I}$ are set homogeneous, corresponding to the case studied par Krook and Wu\cite{Krook76SI,Krook77SI}.
Starting from $f_{_{\bf BKW}}$, the gas relaxes to equilibrium $f_{_{\bf LEQ}}$.

Adimensional time is $\kappa n_I t$ and adimensional distribution function is $\frac{1}{n_I} \left( \sqrt{2\pi} c_{q,I} \right)^3 f$.

\paragraph{For the case in fig. 3b,} we set $\omega = 2\pi$, and $n_I$, $\vec{u}_I$ and $c_{q,I}$ are set heterogeneous to correspond to one state of the undamped oscillation (\emph{breathing mode}\cite{gdo14SI}):

\begin{equation}
\begin{array}{lcl}
n_I &=& p_{{\cal N}(\vec{0}, \frac{c_{q,EQ}^2}{\omega^2})} \vspace{3mm} \\
\vec{u}_I &=& \epsilon \omega \vec{r} \vspace{3mm} \\
c_{q,I} &=& \sqrt{1-\epsilon^2} \ c_{q,EQ}
\end{array}
\end{equation}

with $\epsilon = \frac{\Delta c_{q}^2}{c_{q,EQ}^2}$ where $\Delta c_{q}^2$ is the maximal deviation of $c_{q}^2$ from its equilibrium value $c_{q,EQ}^2$.

With these initial values, and starting from local equilibrium $f_{_{\bf LEQ}}$, the local equilibrium remains unbroken and the gas displays the undamped oscillation for any value of the cross section.
Conversely, starting from out of equilibrium $f_{_{\bf BKW}}$, any positive value of cross section will lead to dampened oscillations.

Adimensional time is $\frac{\omega}{2\pi} t$, adimensional distribution function is $\left( 2\pi \frac{c_{q,EQ}^2}{\omega} \right)^3 f$ and adimensional parameter for cross-section is $\frac{\omega^2}{c_{q,EQ}^3}\kappa$.
 
 Simulation results are given for $\epsilon = 0.2$ and $\frac{\omega^2}{c_{q,EQ}^3}\kappa = 3$.


\subsection{Non-Linear Monte Carlo formulation}


The distribution function obeys the Boltzmann equation:

\begin{equation}
\begin{aligned}
\frac{\partial f}{\partial t} + \ & \vec{c} \cdot \vec{\text{grad}}_{\cal R}(f) + \text{div}_{\cal C}(f \vec{a}) \\
= & \int_{\cal C} d\vec{c}_* \int_{4\pi} d\vec{u} \ \left\Vert \vec{c} - \vec{c}_* \right\Vert \ \sigma(\vec{c},\vec{c}_*)
\left( - f f^* + f^{\alpha} f^{\beta} \right)
\end{aligned}
\end{equation}

where $f \equiv f(\vec{r},\vec{c},t)$, $f^* \equiv f(\vec{r},\vec{c}_*,t)$, $f^{\alpha} \equiv f(\vec{r},\vec{c}_{\alpha},t)$ and $f^{\beta} \equiv f(\vec{r},\vec{c}_{\beta},t)$. 

\bigskip

Velocities $\vec{c}_{\alpha}$ and $\vec{c}_{\beta}$ are functions of $\vec{c}$, $\vec{c}_*$ and $\vec{u}$~: $\vec{c}_{\alpha} = \frac{1}{2}(\vec{c} + \vec{c}_* + \left\Vert \vec{c} - \vec{c}_* \right\Vert\vec{u})$,
$\vec{c}_{\beta} = \frac{1}{2}(\vec{c} + \vec{c}_* - \left\Vert \vec{c} - \vec{c}_* \right\Vert\vec{u})$.

\bigskip
The usual PDE expression of this model, given above, can be translated into its Fredholm counterpart, following:

\begin{equation}
\begin{aligned}
{f(\vec{r},\vec{c},t)} = \ & \int_{-\infty}^t dt' \ \hat{\nu}(t') \ \exp \left( - \int_{t'}^t dt'' \ \hat{\nu}(t'') \right) \times \\
&  \Bigg[ {\cal H}(t_I-t')\; {f(\vec{r}_b(t_I),\vec{c}_b(t_I),t_I)} \\ 
&+ {\cal H}(t'-t_I) \left( 1 - \frac{{\nu(t')}}{\hat{\nu}(t')} \right) {f\left(\vec{r}_b(t'),\vec{c}_b(t'),t'\right)} + \frac{{s(t')}}{\hat{\nu}(t')}\Bigg]
\end{aligned}
\label{eqf}
\end{equation}

with

\begin{equation}
\begin{aligned}
\nu(t') = \int_{\cal C} d\vec{c}_* \int_{4\pi} d\vec{u} & \left\Vert \vec{c}_b(t') - \vec{c}_* \right\Vert \sigma(\vec{c}_b(t'),\vec{c}_*) {f(\vec{r}_b(t'),\vec{c}_*,t')} \\
s(t') = \int_{\cal C} d\vec{c}_* \int_{4\pi} d\vec{u} & \left\Vert \vec{c}_b(t') - \vec{c}_* \right\Vert \sigma(\vec{c}_b(t'),\vec{c}_*) {f(\vec{r}_b(t'),\vec{c}_{\alpha}(t'),t')} {f(\vec{r}_b(t'),\vec{c}_{\beta}(t'),t')}
\end{aligned}
\end{equation}

where $\vec{r}_b$ and $\vec{c}_b$ are location and velocity corresponding to the ballistic path through $\vec{r}$ at time $t$ with velocity $\vec{c}$, such as:
\begin{equation}
\left\{ \begin{array}{l}
\partial_{t'} \vec{r}_b(t') = \vec{c}_b(t') \\
\partial_{t'} \vec{c}_b(t') = \vec{a}(\vec{r}_b(t'))
\end{array} \right.
\nonumber
\end{equation}
and $\vec{r}_b(t)=\vec{r}$, $\vec{c}_b(t)=\vec{c}$.

The product ${f(\vec{r}_b(t'),\vec{c}_{\alpha}(t'),t')} {f(\vec{r}_b(t'),\vec{c_{\beta}}(t'),t')}$ has been treated following the NLMC expansion exposed in Methods.

The exponential term is handled using null collisions technique, as exposed in SI4.
The upper bound for cross section is set at the value it takes at the center of the gas cloud at time of maximal contraction of the undamped oscillation :
\begin{equation}
  \hat{\nu} = \frac{\omega^3}{\left( 2\pi c_{q,EQ}^2 (1- \epsilon) \right)^{3/2}} \ \kappa
  \nonumber
\end{equation}

\subsection{Algorithm}

According to eq. \ref{eqf}, three random variables are defined for the Monte Carlo estimates: $T'$ for $t'$, $\vec{U}$ for $\vec{u}$ and $\vec{C}_*$ for $\vec{c}_*$, with:

\begin{equation}
\begin{array}{lcll}
p_{T'}(t') &=& \hat{\nu} \ \exp \left( - \hat{\nu}(t-t') \right) \; &\text{over} \; ]-\infty, t]\\
p_{\vec{U}}(\vec{u}) &=& \frac{1}{4\pi} \ &\text{over the unit sphere} \\
p_{\vec{C^*}}(\vec{c}_*) &=& p_{{\cal N}(\vec{u}_{Boltz}, c_{q,Boltz}^2)} \ &\text{over velocity space} 
\end{array}
\end{equation}

where $\vec{u}_{Boltz}$ et $c_{q,Boltz}$ are respectively the mean velocity and the mean square speed at the considered time, and set to the values predicted in the case of undamped oscillation.

\noindent ${f(\vec{r},\vec{c},t)}$ can be estimated by this recursive algorithm:
\begin{description}
\item[Initialisation] Sample a date $t'$. 
\item[Recursion termination:]
If $t' \leqslant t_I$ return ${f(\vec{x}_I,\vec{c}_I,t_I)}$.
\item[Recursion] 
$\ $\\
Sample a velocity $\vec{c}_*$ and a unit vector $\vec{u}$. 

Compute the ballistic solution $\vec{x}_b(t')$ and $\vec{c}_b(t')$.

Estimate ${\sf {f}_{*}} \gets {f(\vec{x}_b(t'),\vec{c}_*,t')}$.

Set
\begin{equation}
\begin{array}{lcl}
{\sf n} &\gets& 
{\left( \left\Vert \vec{c}_b(t') - \vec{c}_* \right\Vert \ \sigma(\vec{c}_b(t'),\vec{c}_*,\vec{u})\  {\sf {f}_{*}} \right)}  
\Big/
{\left( p_{\vec{C}_*}(\vec{c}_*)p_{\vec{U}}(\vec{u})\right)} \vspace{3mm} \\ 
{\sf Q} &\gets& {\sf n} / {\hat{\nu}}
\end{array}
\end{equation}

If ${\sf Q} \in [0,1]$, then set ${\sf P} \gets {\sf Q}$ else set ${\sf P} \gets \frac{{\sf Q}}{2{\sf Q}-1}$.

Sample a standard uniform $r$.

If $r > {\sf P}$, then estimate ${\sf {f}_{b}} \gets {f(\vec{x}_b(t'),\vec{c}_b(t'),t')}$  and return $\frac{1-{\sf Q}}{1-{\sf P}} \ {\sf {f}_{b}}$,

Else set
\begin{equation}
\begin{array}{lcl}
\vec{c}_{\alpha}(t') &\gets& \frac{1}{2}(\vec{c}_b(t') + \vec{c}_* + \left\Vert \vec{c}_b(t') - \vec{c}_* \right\Vert\vec{u})\\
\vec{c}_{\beta}(t') &\gets& \frac{1}{2}(\vec{c_b(t')} + \vec{c}_* - \left\Vert \vec{c}_b(t') - \vec{c}_* \right\Vert\vec{u})
\end{array}
\end{equation}

Estimate ${\sf {f}}_{\alpha} \gets {f(\vec{x}_b(t'),\vec{c}_{\alpha}(t'),t')}$.

Estimate ${\sf {f}}_{\beta} \gets {f(\vec{x}_b(t'),\vec{c}_{\beta}(t'),t')}$.

Return $\left({\sf Q} / {\sf P}\right) \left( {{\sf {f}}_{\alpha}\ {\sf {f}}_{\beta}} / {\sf {f}_{*}} \right)$

\end{description}

\clearpage
\pagestyle{empty}
\section{{Extended Data Figures}}
$\;$

\noindent

\hspace{-2.8cm}\includegraphics[width=17cm]{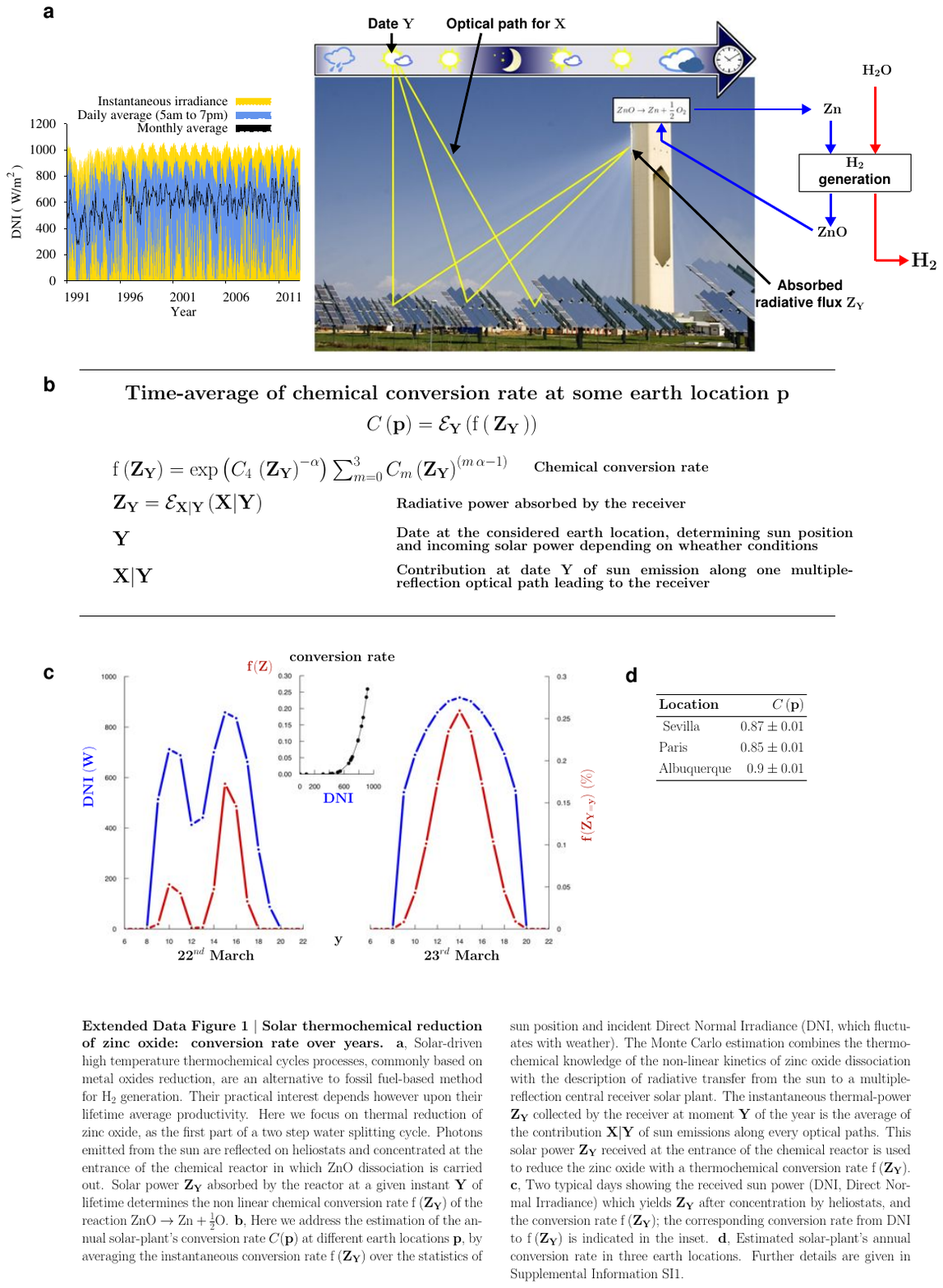}

\hspace{-2.8cm}\includegraphics[width=17cm]{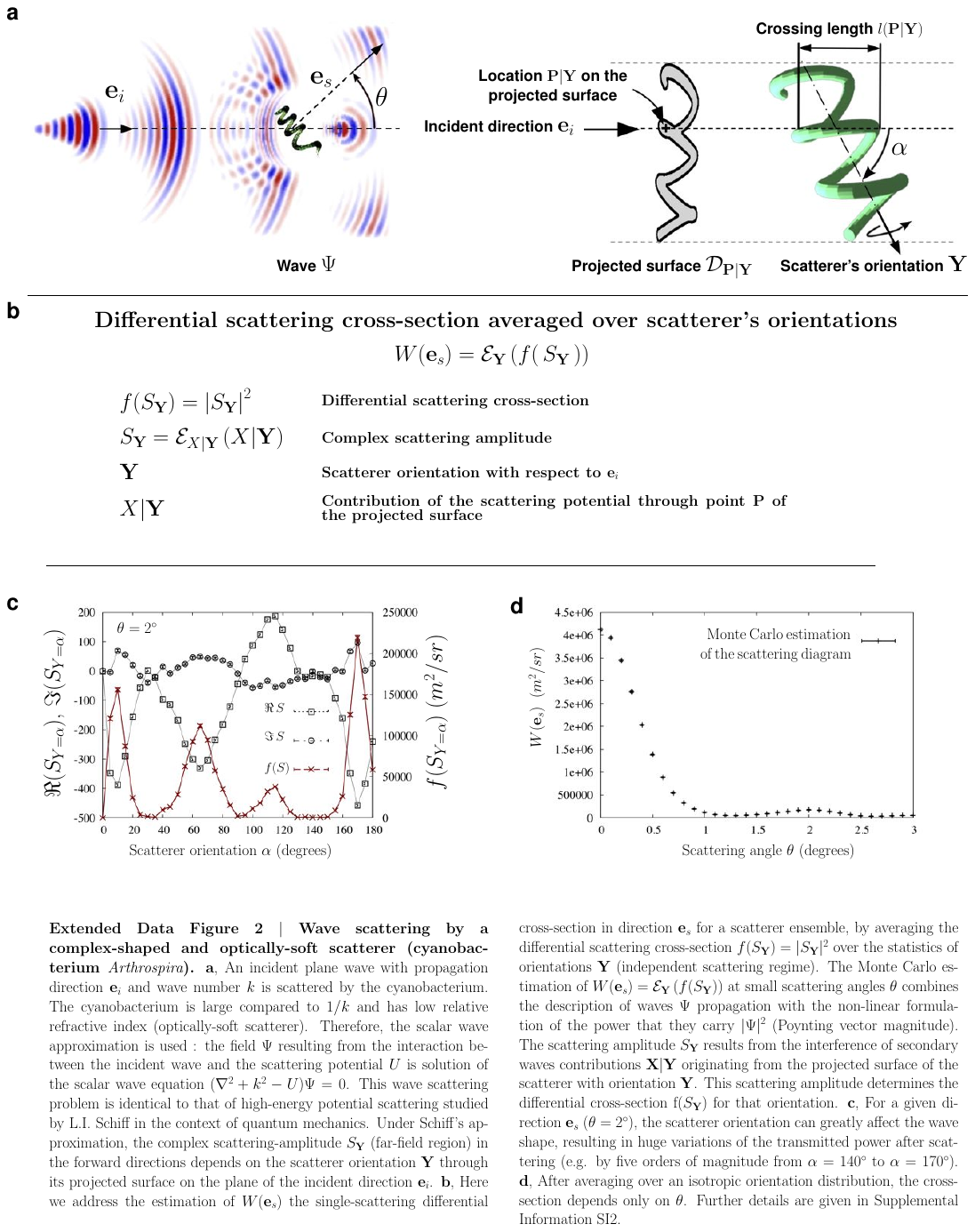}

\hspace{-2.8cm}\includegraphics[width=17cm]{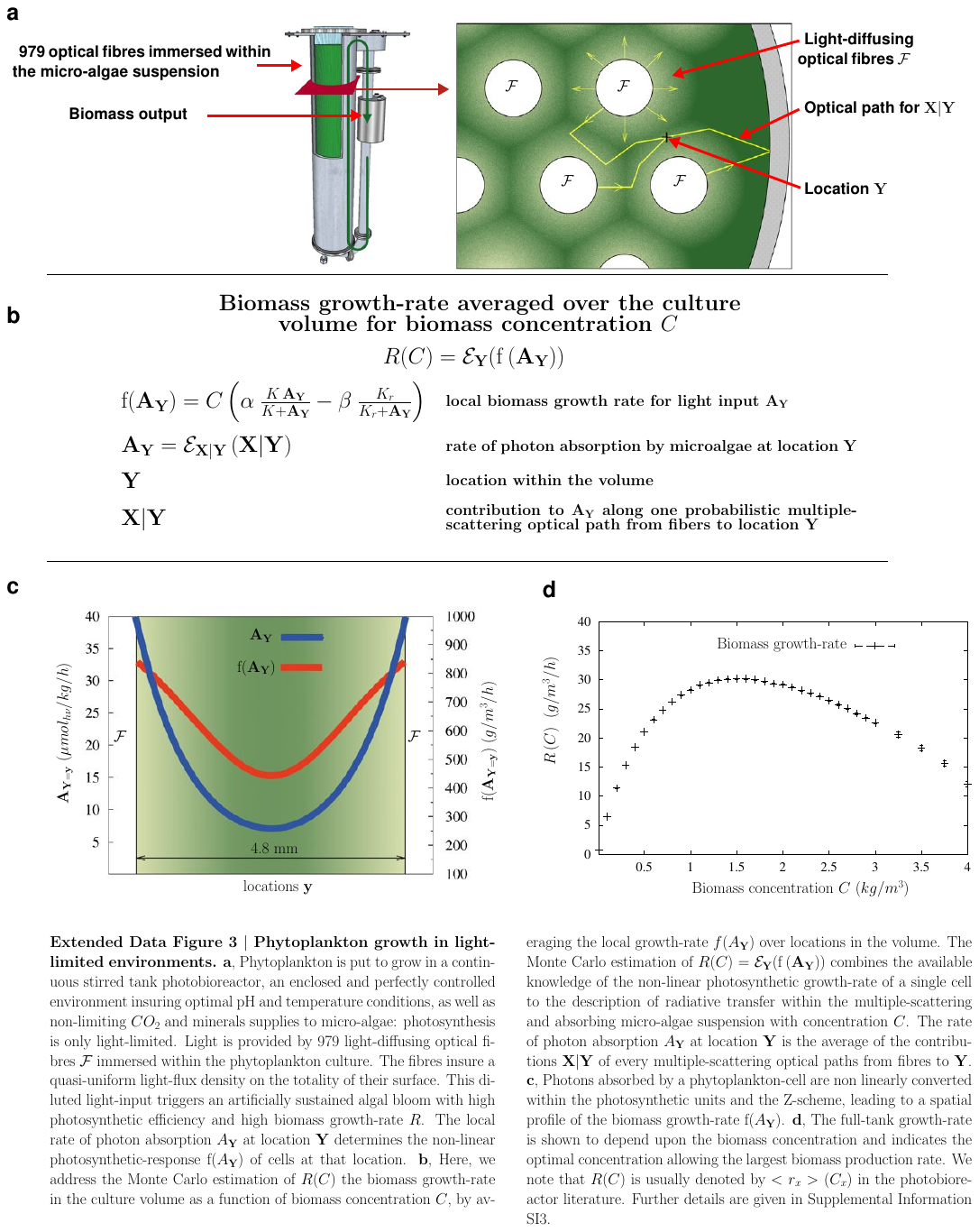}

\hspace{-2.8cm}\includegraphics[width=17cm]{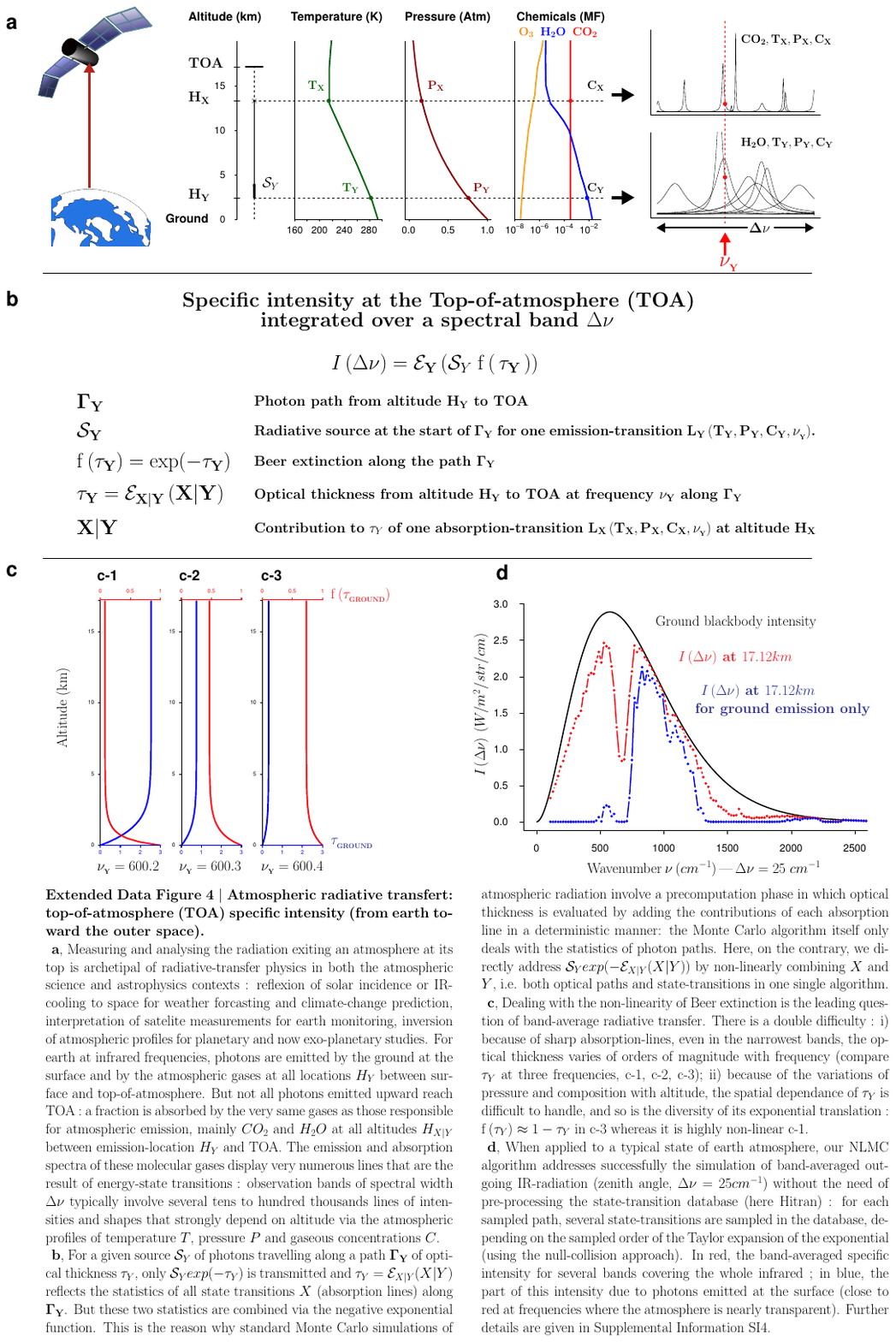}

\end{document}